\documentclass[journal,letterpaper]{IEEEtran}
\usepackage{amsmath,amsfonts}
\usepackage{algorithm}
\usepackage[noEnd=false,indLines=true]{algpseudocodex}
\usepackage{array}
\usepackage[caption=false,font=normalsize,labelfont=sf,textfont=sf]{subfig}
\usepackage{textcomp}
\usepackage{stfloats}
\usepackage{url}
\usepackage{verbatim}
\usepackage{graphicx}
\usepackage{cite}
\usepackage{float}
\usepackage{amssymb} 
\usepackage{xcolor}
\usepackage{amsthm}
\usepackage{booktabs}      
\usepackage{multirow}      
\usepackage{amssymb}       
\usepackage{pifont}        
\usepackage{booktabs}    
\usepackage{pifont}      
\usepackage{array}       

\newcommand{\cmark}{\ding{51}} 
\newcommand{\xmark}{\ding{55}} 

\algrenewcommand{\algorithmiccomment}[1]{\hfill \textit{#1}}

\theoremstyle{remark}
\newtheorem{remark}{Remark}

\usepackage{pifont}

\usepackage{cite}        
\usepackage{hyperref}    

\begin{document}

\title{Secure Energy-Efficient Uplink Transmission in Movable-Element RIS-aided Systems with Movable Antennas and Artificial Noise}

\author{Ayda Nodel Hokmabadi, Mohamed Elhattab, and Chadi Assi \textit{~}}

\maketitle

\begin{abstract}
Secure energy efficiency (SEE) has emerged as a key performance metric for next-generation wireless networks, where energy sustainability and information security must be jointly guaranteed. This paper investigates secure uplink transmission in a full-duplex (FD) base station (BS) system equipped with movable antennas (MAs) and assisted by a movable-element reconfigurable intelligent surface (ME-RIS) in the presence of multiple cooperative passive eavesdroppers. The objective is to maximize SEE by jointly optimizing the users' transmit powers, BS receive postcoders, artificial noise (AN) transmit power and beamforming, RIS phase shifts, and the two-dimensional positions of both the BS antennas and RIS elements. The resulting optimization problem is highly nonconvex due to the fractional SEE objective, coupled secrecy-rate expressions, residual self-interference (SI), unit-modulus RIS phase-shift constraints, movable-position constraints, inter-element spacing requirements, and the nonlinear dependence of the channels on the movable antenna and RIS-element positions. To address these challenges, we propose a hybrid gradient-based meta-learning (H-GML) framework. In the proposed method, the BS receive postcoders and AN direction are updated using closed-form solutions derived from generalized Rayleigh quotient formulations, while the remaining coupled variables are updated by neural meta-optimizers that learn gradient-based update directions directly from the SEE optimization objective without requiring offline labeled training data. Simulation results show that the proposed H-GML design achieves better performance than the AO benchmark and significantly outperforms fixed-geometry, random RIS, no-AN, and no-Eve-knowledge baselines. 
\end{abstract}


\section{Introduction}

The rapid growth in wireless data traffic and the increasing demand for secure applications have made energy-efficient and secure communication essential requirements for sixth-generation (6G) and future wireless networks. Traditional cryptographic approaches, which rely on computational hardness, are becoming less reliable as quantum computing advances and network environments become larger and more complex. This has motivated the investigation of physical-layer security (PLS) techniques, which provide information-theoretic security guarantees independent of computational assumptions~\cite{shannon1949communication}. PLS uses the natural randomness and spatial properties of wireless channels to limit the information that an eavesdropper (Eve) can obtain~\cite{wyner1975wire, mukherjee2014principles}. In a wiretap channel, the achievable secrecy capacity depends on the difference between the channel qualities of the legitimate user and Eve. This motivates the design of signal processing and resource allocation techniques that enhance legitimate links while degrading eavesdropping links.

Movable antennas (MAs) technology has recently emerged as an effective approach to introduce spatial reconfigurability at the base station (BS) side~\cite{zhu2024modeling, ma2025tutorial}. Unlike conventional fixed-position antennas (FPAs), MAs are connected to RF chains through flexible cables or mechanical actuators, allowing each antenna to adjust its position within a bounded region to achieve more favorable channel conditions. The field-response channel model developed in~\cite{zhu2024modeling} provides an accurate characterization of how wireless channels vary with antenna movement, enabling antenna-position optimization. Based on this model, MA-assisted multiuser uplink systems were studied in~\cite{zhu2024multiuser}, where joint optimization of antenna positions, user transmit powers, and BS receive postcoders was shown to significantly reduce the required transmit power while satisfying rate constraints. The secrecy rate of MA-aided wiretap channels was investigated in~\cite{hu2024secure} by jointly optimizing transmit beamforming and antenna positions. This was extended in~\cite{10447471} to jointly minimize power consumption and maximize secrecy rate in a single-user MA system. For scenarios with imperfect eavesdropper channel state information (CSI), robust joint optimization of beamforming and antenna positions was proposed in~\cite{feng2024movable} to guarantee worst-case secrecy performance. In addition, MA-aided secure multiple-input multiple-output (MIMO) communications were studied in~\cite{tang2024secure}, where artificial noise (AN), transmit precoding, and antenna positions were jointly designed. Furthermore, MA-enabled full-duplex secure systems were investigated in~\cite{ding2024secure}, where both uplink and downlink transmissions were protected simultaneously. For the multiuser scenario, secure beamforming with MA arrays was studied in~\cite{cheng2026secure}, showing secrecy-rate gains over FPA-based systems in multiuser wiretap channels.

Despite these advantages, MA technology only reconfigures the transceiver side of the wireless network. Since the propagation environment remains unchanged, the achievable spatial degrees of freedom (DoF) are still limited by the surrounding scattering geometry. Therefore, repositioning the antennas alone cannot fully take advantage of environment-aware communication, especially in scenarios with unfavorable reflections, blockages, or strong Eve channels. In order to overcome these limitations, reconfigurable intelligent surfaces (RISs) have been introduced as a technique to reshape the wireless propagation environment and have recently emerged as a promising technology for next-generation wireless systems. By using large arrays of low-cost passive elements with adjustable phase shifts, RISs can modify the wireless propagation environment without requiring active RF chains, enabling energy-efficient signal enhancement and interference suppression~\cite{wu2019intelligent, basar2019wireless}.

From a PLS perspective, RISs are particularly attractive because their phase-shift design can strengthen the desired signal at legitimate receivers while reducing signal leakage toward Eves. This has motivated extensive research on RIS-assisted secure communication. In~\cite{yu2019enabling}, joint transmit beamforming and RIS phase-shift optimization were proposed to maximize the secrecy rate in a wiretap channel with a single Eve, showing that even passive reflecting surfaces can significantly improve PLS performance. In~\cite{guo2022joint}, both RIS placement and passive beamforming were jointly optimized, highlighting the strong impact of deployment location on secrecy performance. To improve robustness,~\cite{yu2020robust} considered imperfect CSI and AN-aided secure transmission in RIS-assisted systems. In multiuser systems,~\cite{chen2019intelligent} maximized the minimum secrecy rate across multiple legitimate receivers by jointly optimizing BS beamformers and IRS reflection coefficients under both continuous and discrete phase-shift constraints. Additionally,~\cite{hao2022securing} developed an alternating optimization (AO) framework for jointly designing active and passive beamforming to maximize secrecy rates under coupled nonconvex constraints. More recently, secrecy energy efficiency (SEE) in RIS-assisted networks was investigated in~\cite{lu2023secrecy}, where sequential fractional programming and alternating maximization were used to jointly optimize user transmit powers, RIS reflection coefficients, and BS receive filters, achieving notable SEE improvements over conventional designs.

Despite their strong performance gains, conventional RIS architectures assume fixed element positions on a planar surface, which limits the available spatial DoF. To address this limitation, movable-element RIS (ME-RIS) architectures have recently been introduced, allowing each reflecting element to adjust its position within a predefined region in addition to tuning its phase shift~\cite{MERIS5, zhang2024ma_ris_geometry}. This added spatial flexibility enables more effective channel manipulation by jointly optimizing element positions and phase shifts, thereby unlocking spatial DoF that fixed-element RISs cannot access. Such dual reconfigurability allows more flexible control of the wireless environment and can significantly improve system performance~\cite{wei2024movable_ris, li2024mis}. By jointly optimizing element positions and phase shifts, ME-RIS can improve channel gains and provide more flexible interference management.

Prior works have studied ME-RIS with different objectives. Downlink sum-rate maximization in multiuser multiple-input single-output (MISO) networks was studied in~\cite{MERIS2}, while single-user single-input single-output (SISO) settings were considered in~\cite{MERIS3}, typically assuming fixed-position antennas at the BS. Furthermore, secure communication for movable-element-enabled simultaneously transmitting and reflecting surface (ME-STARS)-aided systems with FPA-BS was investigated in~\cite{MERIS1} by maximizing the sum secrecy rate. In addition, element-wise position optimization using successive convex approximation (SCA) was shown in~\cite{MERIS5} to achieve substantial rate gains over conventional RIS designs in both downlink and point-to-point communication scenarios.

Although the movable-antenna-aided BS and ME-RIS each introduce valuable spatial reconfigurability, deploying them separately has limitations for secure and energy-efficient communication. MA-aided systems can only reshape the transceiver-side geometry, leaving the propagation environment unaltered and therefore unable to fully mitigate unfavorable scattering or strong Eve channels. Similarly, ME-RIS improves the propagation environment through element repositioning, but its gains remain bounded when the BS antennas cannot adapt to the optimized reflection geometry. Integrating ME-RIS with an MA-BS overcomes these individual bottlenecks by enabling joint spatial adaptation at both the propagation environment and receiver side. More specifically, ME-RIS reshapes the wireless environment through controllable reflections and flexible element placement, while movable BS antennas adapt the receive array geometry to better capture the resulting signal paths. This additional reconfigurability improves interference suppression, signal alignment, and the channel disparity between legitimate users and Eves, leading to higher secrecy rates and improved SEE in multiuser uplink systems.

On the other hand, full-duplex (FD) communication enables simultaneous uplink reception and downlink transmission over the same time-frequency resource, providing a spectral-efficiency advantage over half-duplex operation. It also allows the BS to transmit artificial noise (AN) to actively degrade the channels of Eves while serving uplink users~\cite{nguyen2018new}. The combination of RIS and FD operation for PLS was investigated in~\cite{guan2022secure}, where joint active beamforming, AN design, and RIS phase-shift optimization were proposed to maximize the sum secrecy rate in a single-user FD system. The results showed that combining RIS-assisted passive beamforming and FD-enabled AN transmission improves the secrecy performance. In addition, the joint MA and ME-RIS design for FD systems, without considering security, was recently studied in~\cite{hokmabadi2026joint}, showing that jointly repositioning BS antennas and RIS elements in a full-duplex MISO network can significantly improve spectral efficiency by suppressing self-interference and enhancing the desired signal strength. However, in practice, residual self-interference (SI), caused by imperfect cancellation of the BS's own transmit signal, must be carefully managed because it can significantly degrade uplink reception quality. Moreover, the combination of ME-RIS and MA for energy-efficient uplink communication was investigated in~\cite{hokmabadi2025joint}, where joint optimization of antenna and element positions, postcoder vectors, and user transmit powers achieved significant energy-efficiency gains over fixed-geometry benchmarks. However, PLS was not considered. In particular, SEE optimization for MA-BS and ME-RIS-assisted systems remains an open research problem.

It is worth mentioning that most existing PLS works on RIS-assisted systems consider relatively simple eavesdropping models, typically involving a single Eve. In more realistic scenarios, multiple Eves may cooperate by sharing their received signals and jointly processing them to decode the legitimate transmissions, resulting in a substantially stronger threat that is not fully addressed by existing designs~\cite{zhang2025secure}. Addressing this cooperative eavesdropping threat in an energy-constrained FD system naturally motivates the maximization of SEE, which jointly accounts for the achievable secrecy rate and the total power consumption. However, the resulting SEE maximization problem is highly nonconvex due to its fractional objective, the coupled legitimate and cooperative-eavesdropping rate expressions, the unit-modulus RIS phase-shift constraints, AN design, residual SI, and the nonlinear dependence of the channels on the positions of the BS antennas and RIS elements. Conventional AO, SCA, and fractional programming (FP) methods typically require repeated convexification and problem-specific solvers. Moreover, their per-iteration complexity grows rapidly with the numbers of movable antennas and RIS elements, making them increasingly impractical for high-dimensional and large-scale movable-element systems.

To address this scalability limitation, meta-learning (meta-L) has recently emerged as an attractive model-based learning paradigm for nonconvex beamforming design~\cite{xia2021meta}. Rather than learning a direct mapping from channel state to beamforming solution, which typically requires large labeled datasets and costly pre-training, meta-L methods train lightweight neural networks to learn an update rule, using the structure of the optimization problem itself as supervision. An early representative approach used long-short-term-memory (LSTM) networks within the inner loop of a meta-L framework to learn a dynamic update strategy for weighted sum-rate maximization in MISO downlink beamforming, outperforming the classical weighted minimum mean square error (WMMSE) algorithm by adapting each variable update to the local geometry of the objective~\cite{xia2021meta}. This line of work was subsequently extended through gradient-based meta-learning (GML), in which the network is fed the optimization gradients directly rather than raw channel information, allowing the learned update rule to generalize across channel realizations without scenario-specific pre-training~\cite{zhu2024robust}. GML-based frameworks have been proposed for joint BS precoding and RIS phase-shift design~\cite{wang2023energy}, for robust RIS-aided beamforming via manifold-constrained meta-L~\cite{zhu2024robust}, and, most recently, for joint precoding and reflection/transmission coefficient design in STAR-RIS-assisted networks, where the GML approach was shown to achieve near-AO performance with substantially reduced runtime and favorable scaling in the number of antennas and RIS elements~\cite{yang2025efficient}.

Meta-L method has also been extended to movable-antenna systems, where the optimization variables include not only beamforming vectors but also continuous antenna positions. A model-driven meta-L framework was used to jointly design BS beamforming, rate-splitting allocation, and MA positions in a CoMP-RSMA system, using separate sub-networks for each variable type to handle their distinct geometries~\cite{amhaz2025enhancing}. A GML-based approach was similarly applied to a movable-antenna-assisted full-duplex RSMA system, jointly optimizing beamforming, MA positions, and power allocation while mitigating self-interference, achieving substantial spectral-efficiency gains over fixed-position-antenna baselines~\cite{khisa2025meta}. Related meta-L and meta-reinforcement-learning frameworks have also been proposed for movable-antenna-aided full-duplex integrated sensing and communication~\cite{amhaz2025meta} and for full-duplex cell-free dual-functional radar-communication systems under carrier frequency offset~\cite{xiu2025metareinforcement}. These works collectively demonstrate that meta-L can effectively handle the joint, highly coupled optimization of beamforming and spatial (antenna or element position) variables without the pre-training overhead of conventional deep-learning approaches, and that it scales more favorably than AO-based methods as the number of movable elements grows.

Despite this progress, existing meta-L frameworks do not jointly consider all design variables in a secrecy-based full-duplex (FD) network with movable degrees of freedom. These variables include MA-BS positioning, ME-RIS positioning and phase-shift design, AN-aided secrecy, and SI management under cooperative multi-Eve threats. To address this gap, this work develops a hybrid meta-L-based SEE optimization framework for ME-RIS-assisted MA-based FD-BS wireless systems with AN-aided secrecy against cooperative Eves. To the best of our knowledge, this joint design has not yet been studied in the meta-L-based beamforming literature.

To better position this work within the existing literature, Table~\ref{tab:comparison} summarizes the key differences between the proposed system and the most closely related works.

\begin{table}[!t]
\centering
\caption{Comparison With Related Works}
\label{tab:comparison}
\renewcommand{\arraystretch}{1.3}
\resizebox{\columnwidth}{!}{%
\begin{tabular}{|c|c|c|c|c|c|c|}
\hline
\textbf{Work} & \textbf{Objective} & \textbf{MA-BS} & 
\textbf{ME-RIS} & \textbf{FD} & 
\textbf{SI mngt} & \textbf{Meta-L} \\
\hline
\cite{MERIS2} & Sum rate           & \xmark & \cmark & \xmark & \xmark & \xmark \\
\hline
\cite{MERIS3} & Achievable rate    & \xmark & \cmark & \xmark & \xmark & \xmark \\
\hline
\cite{MERIS1} & Secrecy rate       & \xmark & \cmark & \xmark & \xmark & \xmark \\
\hline
\cite{MERIS5} & Sum rate      & \xmark & \cmark & \xmark & \xmark & \xmark \\
\hline
\cite{MERIS4} & Sum rate           & \cmark & \cmark & \xmark & \xmark & \xmark \\
\hline
\cite{zhu2024multiuser} & Power minimization & \cmark & \xmark & \xmark & \xmark & \xmark \\
\hline
\cite{ma2025movable} & Secure ISAC        & \cmark & \cmark & \xmark & \xmark & \xmark \\
\hline
\cite{guan2022secure} & Secrecy rate  & \xmark & \xmark & \cmark & \xmark & \xmark \\
\hline
\cite{ding2024secure} & Secrecy rate  & \cmark & \xmark & \cmark & \cmark & \xmark \\
\hline
\cite{yang2025efficient} & Sum rate (STAR-RIS) & \xmark & \cmark & \xmark & \xmark & \cmark \\
\hline
\cite{khisa2025meta} & Sum rate  & \cmark & \xmark & \cmark & \cmark & \cmark \\
\hline
\cite{hokmabadi2025joint} & Energy efficiency & \cmark & \cmark & \xmark & \xmark & \xmark \\
\hline
\cite{hokmabadi2026joint} & Sum rate   & \cmark & \cmark & \cmark & \cmark & \xmark \\
\hline
\textbf{This work} & \textbf{SEE } & \cmark & \cmark & \cmark & \cmark & \cmark \\
\hline
\end{tabular}%
}
\end{table}

\subsection{Contributions}

Motivated by these observations, this paper investigates an ME-RIS-assisted uplink system with movable FD-BS antennas in the presence of multiple cooperative Eves, and proposes a hybrid gradient-based meta-learning (H-GML) framework for SEE maximization. The key insight is that the joint spatial reconfigurability of the MA-BS and ME-RIS elements, together with AN transmission and BS receive postcoding, provides multiple degrees of freedom to enhance the legitimate links, suppress information leakage to the Eves, and reduce SI. Since these variables are highly coupled, they must be optimized jointly. This motivates a hybrid learning-based solution instead of relying only on conventional iterative AO/SCA methods. The main contributions of this paper can be summarized as follows:

\begin{itemize}
\item
We propose an FD uplink system assisted by an ME-RIS and an MA-BS in the presence of multiple cooperative Eves. The system is modeled using the field-response channel framework, where the 2D positions of both BS antennas and RIS elements are treated as optimization variables within predefined feasible regions. To the best of our knowledge, this is the first work to jointly address secrecy and energy efficiency in such a dual-sided movable architecture.

\item
We formulate an SEE maximization problem that jointly optimizes user transmit powers, BS receive postcoder vectors, the AN beamforming direction and power, RIS phase shifts, and the 2D positions of all movable BS antennas and RIS elements. The optimization is performed under per-user quality-of-service (QoS) constraints, user and BS power budgets, unit-modulus RIS phase-shift constraints, movable-region constraints, and minimum inter-element spacing requirements.

\item
To address the highly nonconvex and high-dimensional SEE maximization problem, we propose an H-GML framework. The method is hybrid because it combines closed-form updates for variables with tractable solutions and neural meta-optimizers for the remaining coupled variables. In particular, the BS receive-postcoders are obtained in closed form expressions using the maximum signal-to-interference-plus-noise ratio (SINR) solution derived from a generalized Rayleigh quotient. On the other hand, the AN direction is updated using the principal generalized eigenvector obtained by generalized eigenvalue decomposition. The remaining variables, including user transmit powers, AN power, RIS phase shifts, BS antenna positions, and RIS element positions, are optimized using neural meta-optimizers that learn gradient-based update directions. Different from existing gradient-based meta-learning designs mainly developed for RIS or movable-antenna beamforming~\cite{xia2021meta, wang2023energy, yang2025efficient, khisa2025meta}, the proposed H-GML jointly considers dual-sided spatial mobility, AN-aided secrecy improvement, and SI suppression. Constraints are handled through power reparameterization, phase projection, and penalty terms for QoS, movement-region, and spacing constraints.

\item
We evaluate the proposed framework under different numbers of RIS elements and Monte Carlo channel realizations. The simulations compare the proposed SEE maximization design with several baselines, including energy-efficiency optimization without secrecy, SEE without Eves' channel knowledge, SEE optimization without AN, fixed-geometry schemes, partially movable architectures, and random RIS configurations. The results show the importance of jointly optimizing spatial mobility, RIS phases, AN transmission, and receive postcoding for secure and energy-efficient uplink communication.
\end{itemize}

\subsection{Organization}

The remainder of this paper is organized as follows. Section~II presents the system model. Section~III formulates the SEE maximization problem. Section~IV describes the proposed H-GML algorithm. Section~V provides numerical results, and Section~VI concludes the paper.

\textit{Notation:} In this paper, the boldface lowercase and uppercase letters denote vectors and matrices, respectively. Also, $(\cdot)^T$ and $(\cdot)^H$ represent the transpose and conjugate transpose. $\|\cdot\|$ denotes the Euclidean norm. $\mathbb{C}$ represents the set of complex numbers. $\text{diag}(\cdot)$ is a diagonal matrix from a vector. $\mathcal{CN}(\mu, \sigma^2)$ denotes the circularly symmetric complex Gaussian distribution with mean $\mu$ and variance $\sigma^2$. $\mathbb{E}\{\cdot\}$ denotes expectation.

\section{System Model}

\subsection{Network Model}

\begin{figure}[!t]
    \centering
    \includegraphics[width=0.95\linewidth]{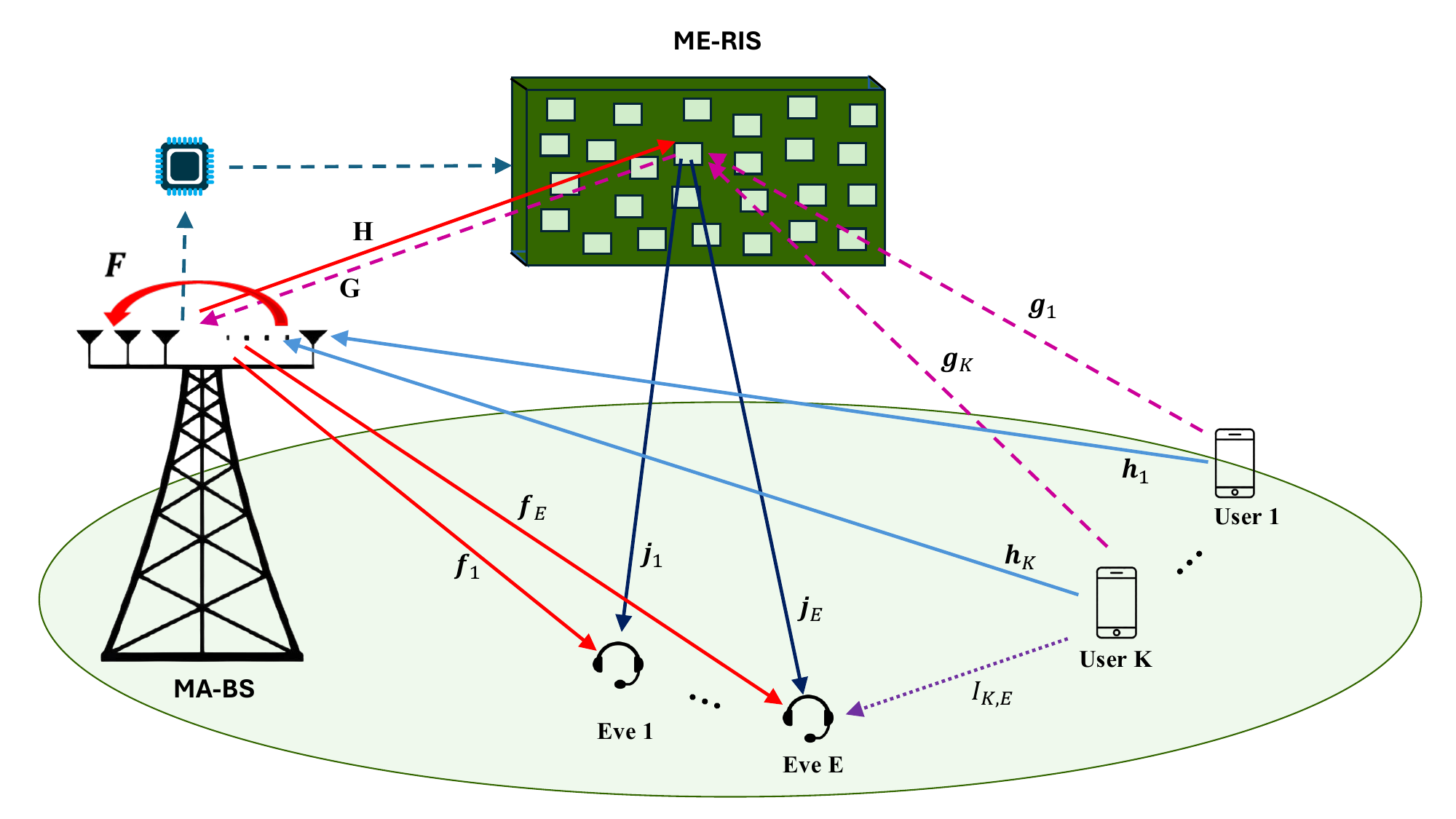}
    \caption{System model for uplink communications}
    \label{fig:system_model}
\end{figure}

As shown in Fig.~\ref{fig:system_model}, we consider an FD uplink communication system assisted by an ME-RIS with $N$ passive reflecting elements. The BS is equipped with $M_r$ movable receive antennas and $M_t$ movable transmit antennas, enabling simultaneous uplink reception and AN transmission. There are $K$ single-antenna legitimate uplink users and $E$ single-antenna Eves. The Eves are assumed to be cooperative. Specifically, although they
remain passive with respect to the legitimate network, they can exchange their received observations through an ideal coordination link and jointly process them to decode the legitimate users' signals. Therefore, the cooperative Eves can be viewed as a distributed multi-antenna Eve, whose combined decoding capability is stronger than that of any individual Eve, and this is called worst-case cooperative eavesdropping, which improves the eavesdropping capability \cite{channel1}.

In addition, the sets of BS transmit antennas, BS receive antennas, RIS elements,
users, and Eves are denoted by $\mathcal{M}_t$,
$\mathcal{M}_r$, $\mathcal{N}$, $\mathcal{K}$, and $\mathcal{E}$,
respectively. Accordingly, each BS receive antenna $m \in \mathcal{M}_r$ is located at a 2D
Cartesian position $\mathbf{u}_m = [x_m^r,\, y_m^r]^T \in
\mathbb{R}^{2}$, and each BS transmit antenna $m \in \mathcal{M}_t$
is located at $\mathbf{t}_m = [x_m^t,\, y_m^t]^T \in \mathbb{R}^{2}$.
Similarly, each RIS element $n \in \mathcal{N}$ is located at
$\mathbf{r}_n = [x_n,\, y_n]^T \in \mathbb{R}^{2}$.
The position matrices are defined as,
\begin{equation}
\begin{aligned}
\mathbf{T}_r &= [\mathbf{u}_1, \ldots, \mathbf{u}_{M_r}]^T \in \mathbb{R}^{M_r \times 2}, \\
\mathbf{T}_t &= [\mathbf{t}_1, \ldots, \mathbf{t}_{M_t}]^T \in \mathbb{R}^{M_t \times 2}, \\
\mathbf{R}   &= [\mathbf{r}_1, \ldots, \mathbf{r}_N]^T \in \mathbb{R}^{N \times 2}.
\end{aligned}
\end{equation}

Furthermore, the RIS applies adjustable phase shifts modeled by the diagonal matrix
\[
  \boldsymbol{\Phi} = \mathrm{diag}\!\left(e^{j\vartheta_1}, \ldots,
  e^{j\vartheta_N}\right), \quad \vartheta_n \in [0,2\pi).
\]

The wireless links in the network are characterized by the following channel vectors and matrices, each of which is a function of the receive and transmit MA, and RIS-element position matrices $\mathbf{T}_r$, $\mathbf{T}_t$, and $\mathbf{R}$ introduced above. The direct uplink channel between user $k$ and the BS receive antennas is denoted by $\mathbf{h}_{k} \in \mathbb{C}^{M_r \times 1}$, while $\mathbf{g}_{k} \in \mathbb{C}^{N \times 1}$ represents the channel between user $k$ and the RIS, and $\mathbf{G} \in \mathbb{C}^{M_r \times N}$ denotes the channel between the RIS and the BS receive antennas. The SI channel between the BS transmit antennas and the BS receive antennas is denoted by $\mathbf{F} \in \mathbb{C}^{M_r \times M_t}$, which shows the FD-induced coupling that the MA positions and the AN beamforming vector $\mathbf{z}$ must jointly mitigate. On the eavesdropping side, $I_{e,k} \in \mathbb{C}$ denotes the direct channel from user $k$ to Eve $e$, $\mathbf{j}_{e} \in \mathbb{C}^{1 \times N}$ denotes the channel between the RIS and Eve $e$, and $\mathbf{f}_{e} \in \mathbb{C}^{1 \times M_t}$ denotes the channel between the BS transmit antennas and Eve $e$. 

\subsection{Signal Model}

Each user $k \in \mathcal{K}$ transmits an information symbol $s_k$
with power $p_k$, where $\mathbb{E}[|s_k|^2] = 1$. Simultaneously,
the BS transmits an AN signal to degrade the Eves'
reception, given by $  \mathbf{x}_{\mathrm{AN}} = \sqrt{p_{\mathrm{an}}}\,\mathbf{z},$
where $\mathbf{z} \in \mathbb{C}^{M_t \times 1}$ is the normalized
AN beamforming vector satisfying $\|\mathbf{z}\|^2 = 1$, and
$p_{\mathrm{an}}$ is the AN transmit power.
Therefore, the received signal at the BS is
\[
  \mathbf{y}_{\mathrm{BS}} =
  \sum_{k=1}^{K} \sqrt{p_k}\,\bar{\mathbf{h}}_{k}\, s_k
  + \sqrt{\eta}\, 
\mathbf{F}\, \mathbf{x}_{\mathrm{AN}}
  + \mathbf{n}_{\mathrm{BS}},
\]
where $\bar{\mathbf{h}}_{k} = \mathbf{h}_{k} +
\mathbf{G}\boldsymbol{\Phi}\mathbf{g}_{k}$ is the effective
uplink channel of user $k$, combining the direct and
RIS-aided paths. The vector
$\mathbf{n}_{\mathrm{BS}} \sim \mathcal{CN}(\mathbf{0},\sigma_u^2
\mathbf{I}_{M_r})$ denotes the additive white Gaussian noise (AWGN)
at the BS, where $\sigma_u^2$ is the noise variance at each BS receive
antenna. The BS applies a linear receive postcoder
$\mathbf{v}_k \in \mathbb{C}^{M_r \times 1}$,
with $\|\mathbf{v}_k\|^2 = 1$, to decode user $k$'s signal,
resulting in $\hat{s}_k = \mathbf{v}_k^H \mathbf{y}_{\mathrm{BS}}$. In addition, the residual SI level is controlled by $\eta \in [0,1]$, where $\eta=0$ corresponds to perfect SI cancellation\cite{hokmabadi2026joint}.
Thus, the SINR for user $k$ at the BS is
\[
  \gamma_k =
  \frac{p_k\,|\mathbf{v}_k^H \bar{\mathbf{h}}_k|^2}
  {\displaystyle\sum_{j \neq k} p_j\,|\mathbf{v}_k^H \bar{\mathbf{h}}_j|^2
   + \eta ~p_{\mathrm{an}}\,|\mathbf{v}_k^H \mathbf{F}\mathbf{z}|^2
   + \sigma_u^2\,\|\mathbf{v}_k\|^2},
\]
and the achievable uplink rate of user $k$ is given by,
\[
  R_k = \log_2(1 + \gamma_k).
\]

Furthermore, the received signal at Eve $e$ when targeting user $k$ is
\[
  y_{e,k} =
  \sum_{k=1}^{K}\sqrt{p_k}\,\tilde{h}_{e,k}\,s_k
  + \mathbf{f}_{e}\mathbf{x}_{\mathrm{AN}} + n_e,
\]
where $\tilde{h}_{e,k} = I_{e,k} + \mathbf{j}_{e}\boldsymbol{\Phi}
\mathbf{g}_{k}$ is the effective channel between the user $k$ and Eve $e$. The scalar noise term
$n_e \sim \mathcal{CN}(0,\sigma_e^2)$ denotes the AWGN at Eve $e$,
where $\sigma_e^2$ is the noise variance at each Eve. The noise terms
$\mathbf{n}_{\mathrm{BS}}$ and ${n_e}$ are assumed to be
mutually independent and independent of all transmitted signals.

Under the worst-case cooperative eavesdropping assumption, each
Eve is assumed to cancel multi-user interference perfectly before decoding the target user's signal~\cite{7222458}. Accordingly, the SINR at the Eve $e$ for decoding user $k$ signal is given by,
\begin{equation}
  \gamma_{e,k} =
  \frac{p_k\,|\tilde{h}_{e,k}|^2}
  {p_{\mathrm{an}}\,|\mathbf{f}_{e}\mathbf{z}|^2 + \sigma_d^2},
\end{equation}

Under cooperative eavesdropping, the $E$ Eves jointly process
their received signals, achieving an equivalent SINR for eavesdropping user $k$ as
\begin{equation}
  \gamma_{E,k} = \sum_{e=1}^{E} \gamma_{e,k}.
\end{equation}
The corresponding eavesdropping rate is then given by
\begin{equation}
  R_{E,k} = \log_2\!\left(1 + \gamma_{E,k}\right).
\end{equation}

This formulation assumes independent noise processes across the cooperative Eves and maximum-ratio combining of their received signals.

\subsection{Secrecy Rate and Secure Energy Efficiency}

The secrecy rate for user $k$ is defined as
\begin{equation}
 R_k^{\mathrm{sec}} = \left[ R_k - R_{E,k} \right]^+,
\end{equation}
where $[x]^+ = \max(x,0)$.
In addition, the sum secrecy rate (SSR) and SEE are given by,
\begin{equation}
  \mathrm{SSR} = \sum_{k=1}^{K} R_k^{\mathrm{sec}}, \quad
\mathrm{SEE} = 
\frac{\mathrm{SSR}}
{\frac{1}{\zeta_p }(\sum_{k=1}^{K} p_k + p_{\mathrm{an}})+ P_c},  
\end{equation}
where $\zeta_p \in (0, 1]$ denotes the power amplifier efficiency and $P_c$ is the circuit power consumption \cite{lu2023secrecy}.

\subsection{Field-Response-Based Channel Model}

All communication links in the considered system are modeled using the
field-response (FR) channel framework~\cite{zhu2024modeling, channel2},
which is particularly suitable for MA-BS and ME-RIS architectures, where the channel response depends explicitly on the positions of antennas and RIS elements. We assume that the movement region of each BS antenna and RIS element
is sufficiently small relative to the corresponding link distances, so
that the far-field plane-wave approximation holds. Under this
assumption, the angles of departure/arrival (AoDs/AoAs) and path gain magnitudes remain approximately constant within the movement region,
while the phase of each multipath component varies with the antenna/element position. This property is the key enabler of channel manipulation through position adaptation. For a generic link with $L$ dominant propagation paths, the channel matrix is given by,
\begin{equation}
  \mathbf{C} = \mathbf{F}_r^H \boldsymbol{\Sigma}\, \mathbf{F}_t,
  \label{eq:generic_channel}
\end{equation}
where $\mathbf{F}_t \in \mathbb{C}^{L \times M}$ and
$\mathbf{F}_r \in \mathbb{C}^{L \times N}$ are the transmit and
receive field-response matrices (FRMs), respectively, and
$\boldsymbol{\Sigma} \in \mathbb{C}^{L \times L}$ is the diagonal path-response matrix (PRM). The field-response vector (FRV) at position $\mathbf{p} = [x, y]^T \in \mathbb{R}^2$ for a wave arriving from direction $(\theta,\phi)$ is given by,
\begin{equation}
  a(\mathbf{p}, \boldsymbol{\psi}) =
  e^{j\frac{2\pi}{\lambda}
  \left( x\sin\theta\cos\phi + y\sin\theta\sin\phi \right)},
\end{equation}
where $\lambda$ is the carrier wavelength and $\boldsymbol{\psi} = (\theta,\phi)$ denotes the AoA/AoD pair. In addition, the FRV at position $\mathbf{p}$ can be defined as
\begin{equation}
  [\mathbf{f}(\mathbf{p})]_\ell =
  e^{j\frac{2\pi}{\lambda}\,\rho_\ell(\mathbf{p})},
  \quad \ell = 1,\ldots,L,
\end{equation}
where $\rho_\ell(\mathbf{p})$ denotes the projected path length onto the 2D movement plane. Moreover, the FRM for an array of $Q$ elements at positions
$\{\mathbf{p}_q\}_{q=1}^{Q}$ is then
\begin{equation}
  \mathbf{F} = \bigl[\mathbf{f}(\mathbf{p}_1),\,\ldots,\,
  \mathbf{f}(\mathbf{p}_Q)\bigr] \in \mathbb{C}^{L \times Q}.
\end{equation}

Using the above framework, the channel matrices in the considered system can be constructed based on the field-response representation.
Specifically, the RIS-BS channel is given as
\begin{equation}
  \mathbf{G}(\mathbf{T}_r, \mathbf{R}) =
  \mathbf{F}_{r,\mathrm{RB}}^H(\mathbf{T}_r)\,
  \boldsymbol{\Sigma}_{\mathrm{RB}}\,
  \mathbf{F}_{t,\mathrm{RB}}(\mathbf{R})
  \in \mathbb{C}^{M_r \times N}.
\end{equation}
On the other hand, the direct uplink channel from user $k$ to the BS is given by
\begin{equation}
  \mathbf{h}_{k}(\mathbf{T}_r) =
  \mathbf{F}_{r,\mathrm{Bu}}^H(\mathbf{T}_r)\,
  \boldsymbol{\sigma}_{\mathrm{Bu},k},
\end{equation}
while the RIS-user channel is modeled as
\begin{equation}
  \mathbf{g}_{k}(\mathbf{R}) =
  \mathbf{F}_{r,\mathrm{Ru}}^H(\mathbf{R})\,
  \boldsymbol{\sigma}_{\mathrm{Ru},k}.
\end{equation}

For the eavesdropping links, the direct user-Eve channel is
denoted by $I_{e,k}$, whereas the RIS-Eve and
BS-Eve channels are represented by $\mathbf{j}_{e}(\mathbf{R})$ and $\mathbf{f}_{e}(\mathbf{T}_t)$, respectively. Finally, due to FD operation, the BS experiences residual SI, which is modeled as
\begin{equation}
  \mathbf{F}(\mathbf{T}_r, \mathbf{T}_t) =
  \mathbf{F}_{r,\mathrm{SI}}^H(\mathbf{T}_r)\,
  \boldsymbol{\Sigma}_{\mathrm{SI}}\,
  \mathbf{F}_{t,\mathrm{SI}}(\mathbf{T}_t).
\end{equation}

\begin{remark}
  The far-field plane-wave assumption holds when the movement
  regions are limited relative to the link distances. Extending
  the model to near-field scenarios constitutes an interesting
  direction for future work.
\end{remark}

\section{Problem Formulation}

Our objective is to improve the PLS of the considered MA-BS and ME-RIS-aided
FD uplink system by jointly optimizing the users' transmit
powers, the BS receive postcoder vectors, the AN power and direction
vector, the ME-RIS phase-shift matrix, and the 2D positions of the
MAs at the BS and elements at the RIS. The study aims to
maximize the SEE, constrained by the maximum transmit power for users
and BS, the QoS constraint on the minimum achieved rate at the BS for
each user's signal, the unit-modulus constraint for RIS phase-shift
elements, the movable-region constraints for the movable antennas
and elements, and minimum inter-element spacing between two movable antennas/elements.

We define the set of optimization variables to simplify the
formulation as
\begin{equation}
  \mathcal{X} = \bigl\{\{p_k\}_{k=1}^{K},\,\{ \mathbf{v}_k \}_{k=1}^{K},\,\mathbf{z},\,
  p_{\mathrm{an}},\,\boldsymbol{\Phi},\,\mathbf{T}_r,\,
  \mathbf{T}_t,\,\mathbf{R}\bigr\},
\end{equation}
where $p_k$ is the transmit power of user $k$; $\mathbf{v}_k \in
\mathbb{C}^{M_r \times 1}$ is the receive postcoder at the BS for
user $k$; $\mathbf{z} \in \mathbb{C}^{M_t \times 1}$ is the AN
beamforming direction vector; $p_{\mathrm{an}}$ is the AN transmit
power; $\boldsymbol{\Phi}$ is the RIS phase-shift matrix; and
$\mathbf{T}_r \!=\! [\mathbf{u}_1,\!\ldots\!,\mathbf{u}_{M_r}]^T$,
$\mathbf{T}_t \!=\! [\mathbf{t}_1,\!\ldots\!,\mathbf{t}_{M_t}]^T$,
and $\mathbf{R} \!=\! [\mathbf{r}_1,\!\ldots\!,\mathbf{r}_N]^T$
collect the 2D positions of the BS receive antennas, BS transmit
antennas, and RIS elements, respectively.

Accordingly, the SEE maximization problem can be formulated as
\begin{align}
  \max_{\mathcal{X}} \quad
  & \mathrm{SEE}
  \tag{P1}\label{P1} \\
  \text{s.t.} \quad
  & R_k \ge R_k^{\mathrm{th}},
    \quad \forall k \in \mathcal{K},
  \tag{C1}\label{C1} \\
  & 0 \le p_k \le P_k^{\max},
    \quad \forall k \in \mathcal{K},
  \tag{C2}\label{C2} \\
  & 0 \le p_{\mathrm{an}} \le P_{\mathrm{an}}^{\max},
  \tag{C3}\label{C3} \\
  & \|\mathbf{v}_k\|_2^2 = 1,
    \quad \forall k \in \mathcal{K},
  \tag{C4}\label{C4} \\
  & \|\mathbf{z}\|_2^2 = 1,
  \tag{C5}\label{C5} \\
  & \left|[\boldsymbol{\Phi}]_{n,n}\right| = 1,
    \quad \forall n \in \mathcal{N},
  \tag{C6}\label{C6} \\
  & \mathbf{u}_m \in \mathcal{U},
    \quad \forall m \in \mathcal{M}_r,
  \tag{C7}\label{C7} \\
  & \mathbf{t}_m \in \mathcal{T},
    \quad \forall m \in \mathcal{M}_t,
  \tag{C8}\label{C8} \\
  & \mathbf{r}_n \in \mathcal{R},
    \quad \forall n \in \mathcal{N},
  \tag{C9}\label{C9} \\
  & \|\mathbf{u}_m - \mathbf{u}_{m'}\|_2 \ge d_0,
    \quad \forall m \neq m'  \in \mathcal{M}_r,
  \tag{C10}\label{C10} \\
  & \|\mathbf{t}_m - \mathbf{t}_{m'}\|_2 \ge d_0,
    \quad \forall m \neq m' \in \mathcal{M}_t,
  \tag{C11}\label{C11} \\
  & \|\mathbf{r}_n - \mathbf{r}_{n'}\|_2 \ge d_0,
    \quad \forall n \neq n' \in \mathcal{N},
  \tag{C12}\label{C12}
\end{align}
where constraint~\eqref{C1} ensures the QoS that guarantees a minimum rate
$R_k^{\mathrm{th}}$ for each user. Constraints~\eqref{C2}-\eqref{C3}
limit the transmit powers of the users and the AN signal.
Constraints~\eqref{C4}-\eqref{C5} denote that the receive postcoders
and the AN direction vector have unit norm. In addition,
constraint~\eqref{C6} sets the unit-modulus constraint on each RIS
phase-shift element. Constraints~\eqref{C7}-\eqref{C9} restrict the
2D positions of the BS receive and transmit antennas and RIS elements
to their respective feasible regions $\mathcal{U}$, $\mathcal{T}$,
and $\mathcal{R}$. Constraints~\eqref{C10}-\eqref{C12} enforce a
minimum inter-element spacing of $d_0$ to prevent strong mutual
coupling among antennas and RIS elements.

Problem~\eqref{P1} is highly non-convex due to the fractional form
of the SEE objective, the non-convex secrecy-rate expressions arising from the $[\cdot]^+$ operator, the cooperative multi-Eve aggregation, the unit-modulus constraint on the RIS phase shifts, and the position-dependent field-response channel structure induced by the movable antennas and RIS elements. The
dimensionality of $\mathcal{X}$ grows with $N$, $M_t$, $M_r$, and $K$, and the channels depend on $\mathbf{T}_t$, $\mathbf{T}_r$, and $\mathbf{R}$ in a highly nonlinear manner through the field-response
vectors. Classical AO frameworks that rely on per-block SCA convexification therefore require repeated linearization, trust-region control, and convex solver calls for the position and phase-shift blocks at every outer iteration, which becomes computationally complex as $N$ and the number of movable antennas increase. To address this scalability bottleneck while retaining the
structural advantages of closed-form updates where they are available, we propose an H-GML
framework, which decomposes $\mathcal{X}$ into (i) variables that admit closed-form optimal updates given the remaining variables, and
(ii) coupled, constrained variables that are updated using
lightweight neural meta-optimizers trained directly on the
optimization gradients of problem~\eqref{P1}.

\section{Proposed H-GML Algorithm}

In this section, we develop an H-GML framework to solve
problem~\eqref{P1}. The BS receive postcoders and the AN direction are
updated in closed form, while the user transmit powers $\{p_k\}$, AN
power $p_{\rm an}$, RIS phase vector $\boldsymbol{\theta}$, and movable
positions $\mathbf{T}_t$, $\mathbf{T}_r$, and $\mathbf{R}$ are updated
by neural meta-optimizers. The proposed meta-optimizers are trained
online for each channel realization from the gradients of the objective
function, without requiring offline labeled datasets.

\subsection{Closed-Form Update of Receive Postcoders}

For fixed remaining variables, the postcoder $\mathbf{v}_k$ maximizes
the SINR of user $k$. Defining
\begin{equation}
\mathbf{R}_k =
\sigma_u^2 \mathbf{I}
+\eta \, p_{\rm an}(\mathbf{F}\mathbf{z})(\mathbf{F}\mathbf{z})^H
+\sum_{j\neq k}p_j\bar{\mathbf{h}}_j\bar{\mathbf{h}}_j^H,
\end{equation}
the corresponding generalized Rayleigh quotient yields
\begin{equation}
\mathbf{v}_k^{\star}
=
\frac{\mathbf{R}_k^{-1}\bar{\mathbf{h}}_k}
{\left\|\mathbf{R}_k^{-1}\bar{\mathbf{h}}_k\right\|_2}.
\label{eq:v_closed_form}
\end{equation}
This update is applied at every inner iteration of the proposed
algorithm.

\subsection{Closed-Form Update of AN Direction}

For fixed remaining variables, the AN direction $\mathbf{z}$ is designed
to strengthen the AN leakage toward the cooperative Eves while reducing
its leakage to the BS receivers. This leads to
\begin{equation}
\max_{\|\mathbf{z}\|_2=1}
\frac{\mathbf{z}^H\mathbf{A}\mathbf{z}}
{\mathbf{z}^H\mathbf{B}\mathbf{z}},
\label{eq:z_subproblem}
\end{equation}
where
\begin{equation}
\mathbf{A}=\sum_{e=1}^{E}\mathbf{f}_e^H\mathbf{f}_e,
\qquad
\mathbf{B}=\sum_{k=1}^{K}\mathbf{F}^H
\mathbf{v}_k\mathbf{v}_k^H\mathbf{F}.
\end{equation}
Thus, $\mathbf{z}^{\star}$ is obtained as the dominant generalized
eigenvector of the matrix pair $(\mathbf{A},\mathbf{B})$.

\subsection{GML-Based Update of the Remaining Variables}

The remaining variables are highly coupled and do not admit tractable
closed-form solutions. Therefore, we use lightweight feed-forward
meta-optimizers to map normalized gradients into update directions. The
power constraints are enforced by the reparameterization
\begin{align}
p_k &= P_k^{\max}\sigma(a_k), \quad k\in\mathcal{K}, \label{eq:power_reparam_user}\\
p_{\rm an} &= P_{\rm an}^{\max}\sigma(a_{\rm an}), \label{eq:power_reparam_an}
\end{align}
where $\sigma(\cdot)$ is the logistic sigmoid function. The RIS
unit-modulus constraint is handled through the phase retraction
\begin{equation}
\theta_n \leftarrow \theta_n \bmod 2\pi,\quad n=1,\ldots,N.
\label{eq:phase_retraction}
\end{equation}

Let
\begin{equation}
\mathcal{S}\triangleq\{\mathbf{T}_t,\mathbf{T}_r,\mathbf{R}\}.
\end{equation}
The constraint-violation terms are defined as
\begingroup
\small
\setlength{\jot}{2pt}
\begin{align}
\mathcal{V}_1
&=
\sum_{k=1}^{K}\left[R_k^{\rm th}-R_k\right]_+^2,
\label{eq:v1}\\
\mathcal{V}_2(\mathbf{X})
&=
\sum_{m<m'}
\left[d_0-\|\mathbf{x}_m-\mathbf{x}_{m'}\|_2\right]_+^2,
\quad \mathbf{X}\in\mathcal{S},
\label{eq:v2}\\
\mathcal{V}_3(\mathbf{X})
&=
\left\|[\mathbf{X}_0-\rho-\mathbf{X}]_+\right\|_F^2
+
\left\|[\mathbf{X}-\mathbf{X}_0-\rho]_+\right\|_F^2,
\nonumber\\
&\quad \mathbf{X}\in\mathcal{S}.
\label{eq:v3}
\end{align}
\endgroup
where $\mathcal{V}_1$, $\mathcal{V}_2(\mathbf{X})$, and
$\mathcal{V}_3(\mathbf{X})$ correspond to the QoS, minimum-spacing, and
movement-region violations, respectively.

For the secrecy-aware design, we set $\mathcal{J}=\mathrm{SEE}$ and
$\mathcal{A}=\mathrm{SSR}$. For the EE baselines, $\mathcal{J}$ and
$\mathcal{A}$ are replaced by EE and the sum rate, respectively. The
losses used for the power, RIS phase, movable-position, and outer
meta-training updates are respectively defined as
\begingroup
\small
\setlength{\jot}{2pt}
\begin{align}
\mathcal{L}_{\rm pow}
&=
-\mathcal{J}
+
\mu_1\mathcal{V}_1,
\label{eq:loss_power}\\
\mathcal{L}_{\boldsymbol{\theta}}
&=
-\mathcal{A}
+
\mu_1\mathcal{V}_1,
\label{eq:loss_theta}\\
\mathcal{L}_{\mathbf{X}}
&=
-\mathcal{A}
+
\mu_1\mathcal{V}_1
+
\mu_2\mathcal{V}_2(\mathbf{X})
+
\mu_3\mathcal{V}_3(\mathbf{X}),
\quad \mathbf{X}\in\mathcal{S},
\label{eq:loss_position}\\
\mathcal{L}_{out,o}
&=
-\mathcal{J}(\mathcal{X}_o^{\rm new})
+
\mu_1\mathcal{V}_1
+
\mu_2\sum_{\mathbf{X}\in\mathcal{S}}\mathcal{V}_2(\mathbf{X})
\nonumber\\
&\quad+
\mu_3\sum_{\mathbf{X}\in\mathcal{S}}\mathcal{V}_3(\mathbf{X}).
\label{eq:outer_loss}
\end{align}
\endgroup
Here, $\mu_1$, $\mu_2$, and $\mu_3$ are fixed penalty weights. For each
active block $\mathbf{x}$, the corresponding normalized gradient is
\begin{equation}
\mathbf{g}_{\mathbf{x}}
=
\frac{\nabla_{\mathbf{x}}\mathcal{L}_{\mathbf{x}}}
{\|\nabla_{\mathbf{x}}\mathcal{L}_{\mathbf{x}}\|_2+\epsilon},
\label{eq:normalized_gradient}
\end{equation}
where $\mathcal{L}_{\mathbf{x}}$ is selected from
\eqref{eq:loss_power}--\eqref{eq:loss_position} according to the
variable block. The meta-update is then written as
\begin{equation}
\mathbf{x}
\leftarrow
\Pi_{\mathbf{x}}
\left(
\mathbf{x}
+
f_{\boldsymbol{\omega}_{\mathbf{x}}}
(\mathbf{g}_{\mathbf{x}})
\right),
\label{eq:meta_update}
\end{equation}
where $\Pi_{\mathbf{x}}(\cdot)$ is the phase retraction for
$\boldsymbol{\theta}$ and the identity mapping for the remaining raw
variables.

The meta-optimizer parameters are collected as
\begin{equation}
\boldsymbol{\omega}
=
\{
\boldsymbol{\omega}_p,
\boldsymbol{\omega}_{\rm an},
\boldsymbol{\omega}_{\theta},
\boldsymbol{\omega}_{R},
\boldsymbol{\omega}_{T_r},
\boldsymbol{\omega}_{T_t}
\}.
\end{equation}
They are trained online using inner, outer, and epoch iterations. The
inner loop updates the optimization variables using
\eqref{eq:meta_update}. The outer loop evaluates multiple trajectories,
where the first one starts from the current solution and the others are
randomly initialized. After all outer trajectories are evaluated, the
average outer loss
\begin{equation}
\bar{\mathcal{L}}
=
\frac{1}{N_o}
\sum_{o=1}^{N_o}
\mathcal{L}_{out,o}
\label{eq:average_outer_loss}
\end{equation}
is used to update $\boldsymbol{\omega}$ via Adam. The best trajectory is
carried forward to the next epoch, and the best solution over all epochs
is returned.

\begin{figure}
    \centering
    \includegraphics[width=0.9\linewidth]{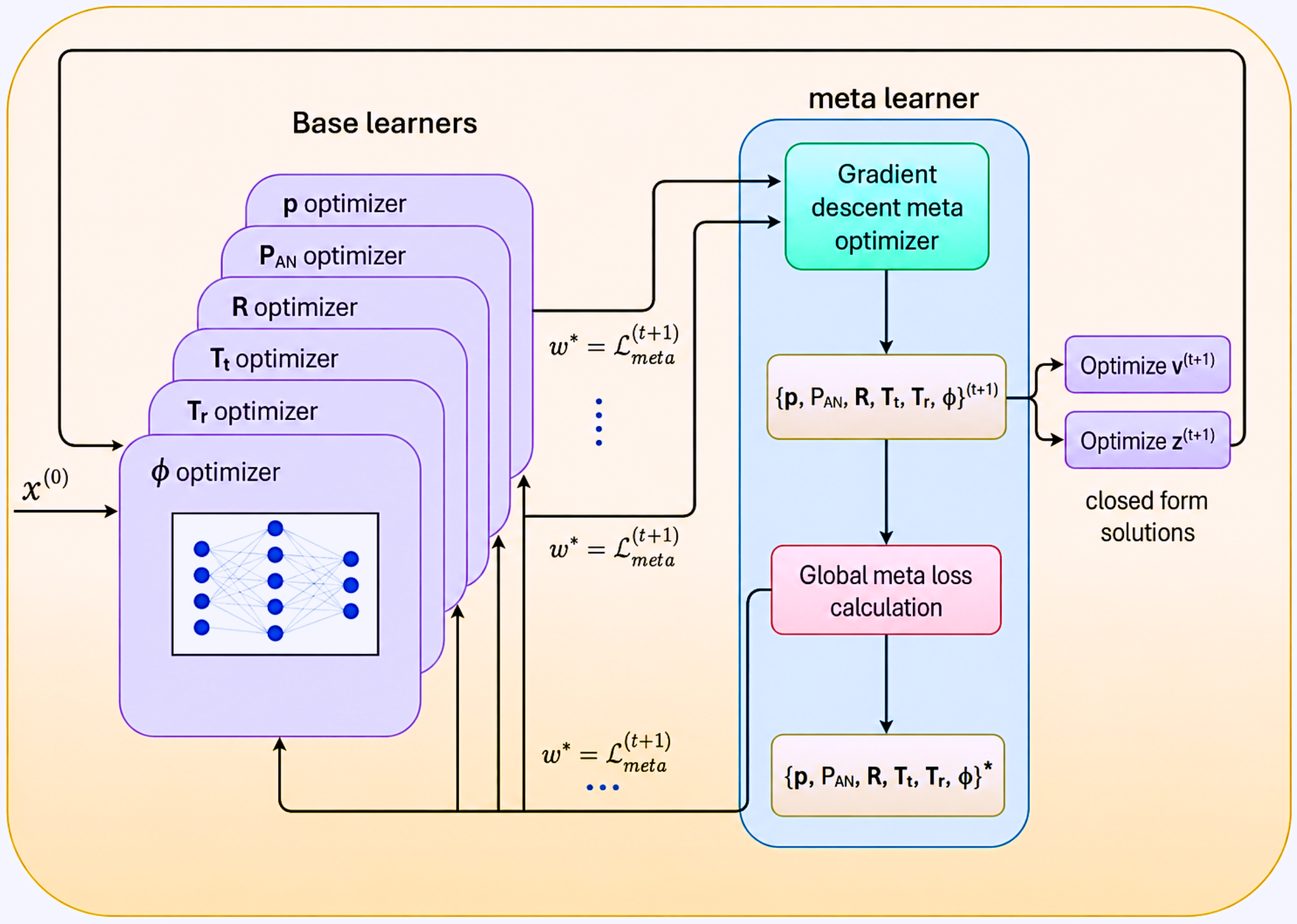}
    \caption{Proposed H-GML optimization algorithm structure.}
    \label{fig:algorithm}
\end{figure}

Given these considerations, the high-level structure of the proposed
H-GML optimization algorithm is shown in Fig.~\ref{fig:algorithm}, while
its detailed operation is summarized in Algorithm~\ref{alg:hgml}.

\begin{algorithm}
\caption{H-GML Optimization Algorithm}
\label{alg:hgml}
\begin{algorithmic}[1]
\setlength{\itemsep}{1.2pt}

    \State Initialize $\mathbf{a}_p^{(0)}$, $a_{\rm an}^{(0)}$,
    $\boldsymbol{\theta}^{(0)}$, $\mathbf{R}^{(0)}$,
    $\mathbf{T}_r^{(0)}$, and $\mathbf{T}_t^{(0)}$
    \State Apply \eqref{eq:phase_retraction} and recover
    $p_k^{(0)}$ and $p_{\rm an}^{(0)}$ from
    \eqref{eq:power_reparam_user} and \eqref{eq:power_reparam_an}
    \State Compute $\mathbf{v}_k^{(0)}$, $\forall k$, and
    $\mathbf{z}^{(0)}$ using the closed-form updates
    \State Initialize $\boldsymbol{\omega}^{(0)}$,
    $\mathcal{X}^{[0]}=\mathcal{X}^{(0)}$,
    $\mathcal{J}^{\star}=0$, and
    $\mathcal{X}^{\star}=\mathcal{X}^{[0]}$

    \For{$t=1,\ldots,T_{\max}$}
        \State Set $\bar{\mathcal{L}}=0$ and $\mathcal{J}_t^{\max}=0$

        \For{$o=1,\ldots,N_o$}
            \State Initialize $\mathcal{X}^{(0,o)}$ from
            $\mathcal{X}^{[t-1]}$ if $o=1$; otherwise randomly initialize it

            \For{$i=1,\ldots,N_i$}
                \State Update $\mathbf{v}_k^{(i-1,o)}$ using
                \eqref{eq:v_closed_form}, $\forall k$
                \State Update $\mathbf{z}^{(i-1,o)}$ using
                \eqref{eq:z_subproblem} if AN is enabled

                \State Compute $\mathcal{L}_{\rm pow}^{(i-1,o)}$,
                $\mathcal{L}_{\boldsymbol{\theta}}^{(i-1,o)}$,
                $\mathcal{L}_{\mathbf{X}}^{(i-1,o)}$, $\mathbf{X}\in\mathcal{S}$

                \State Compute the normalized gradients using
                \eqref{eq:normalized_gradient}

                \State Update $\mathbf{a}_p^{(i,o)}$ and
                $a_{\rm an}^{(i,o)}$ using their meta-optimizers
                \State Update $\boldsymbol{\theta}^{(i,o)}$ using
                $f_{\boldsymbol{\omega}_{\theta}}(\cdot)$ and
                \eqref{eq:phase_retraction}
                \State Update each $\mathbf{X}^{(i,o)}$, $\mathbf{X}\in\mathcal{S}$,
                using its corresponding meta-optimizer

                \State Recover $p_k^{(i,o)}$ and $p_{\rm an}^{(i,o)}$
                and recompute the channels
            \EndFor

            \State Set $\mathcal{X}_o^{\rm new}=\mathcal{X}^{(N_i,o)}$
            \State Recompute $\mathbf{v}_k^{\rm new,o}$, $\forall k$,
            and $\mathbf{z}^{\rm new,o}$
            \State Evaluate $\mathcal{J}^{o}=\mathcal{J}(\mathcal{X}_o^{\rm new})$
            and compute $\mathcal{L}_{out,o}$ using \eqref{eq:outer_loss}
            \State $\bar{\mathcal{L}}\leftarrow
            \bar{\mathcal{L}}+\mathcal{L}_{out,o}$

            \If{$\mathcal{J}^{o}>\mathcal{J}^{\star}$}
                \State $\mathcal{J}^{\star}\leftarrow\mathcal{J}^{o}$,
                $\mathcal{X}^{\star}\leftarrow\mathcal{X}_o^{\rm new}$
            \EndIf
            \If{$\mathcal{J}^{o}>\mathcal{J}_{t}^{\max}$}
                \State $\mathcal{J}_{t}^{\max}\leftarrow\mathcal{J}^{o}$,
                $\mathcal{X}^{[t]}\leftarrow\mathcal{X}_o^{\rm new}$
            \EndIf
        \EndFor

        \State $\bar{\mathcal{L}}\leftarrow \bar{\mathcal{L}}/N_o$
        \State Update $\boldsymbol{\omega}^{(t)}$ by applying Adam to
        $\nabla_{\boldsymbol{\omega}}\bar{\mathcal{L}}$
    \EndFor

    \State \Return $\mathcal{X}^{\star}$
\end{algorithmic}
\end{algorithm}

Unlike conventional AO methods that solve each block through
problem-specific approximations, the proposed H-GML algorithm updates
the coupled power, phase, and position variables through learned
gradient-based steps, while only the receive postcoders and AN direction
are updated in closed form. This avoids repeated SCA-based subproblem
solving and reduces the computational burden as the number of RIS
elements increase.

\subsection{Computational Complexity Analysis}

We analyze the per-epoch computational complexity of the proposed H-GML
algorithm. Each epoch contains $N_o$ outer trajectories, and each
trajectory performs $N_i$ inner iterations.

At each inner iteration, the main operations are the closed-form
postcoder update, the closed-form AN-direction update, channel
recomputation, meta-optimizer inference, and constraint-violation
evaluation. The receive postcoder update requires solving one linear
system of size $M_r$ for each user, which has complexity
$\mathcal{O}(K M_r^3)$~\cite{golub}. The AN direction is obtained from
a generalized eigenvalue decomposition of an $M_t\times M_t$ matrix
pair, with complexity $\mathcal{O}(M_t^3)$~\cite{golub}.

Using the field-response channel model with $L$ propagation paths, the
channel recomputation cost is
\begin{equation}
\mathcal{C}_{\rm ch}
=
\mathcal{O}\!\left(
L(M_rN+KN+EN+M_rM_t+EM_t+KM_r)
\right).
\end{equation}
The six meta-optimizers associated with
$\mathbf{a}_p$, $a_{\rm an}$, $\boldsymbol{\theta}$, $\mathbf{R}$,
$\mathbf{T}_r$, and $\mathbf{T}_t$ have total inference cost
\begin{equation}
\mathcal{C}_{\rm NN}
=
\mathcal{O}\!\left(
H(K+1+3N+2M_r+2M_t)
\right),
\end{equation}
where $H$ is the hidden-layer width. The evaluation of the constraint
violations in \eqref{eq:v1}--\eqref{eq:v3} is dominated by the
pairwise spacing terms, yielding
\begin{equation}
\mathcal{C}_{\rm v}
=
\mathcal{O}\!\left(N^2+M_r^2+M_t^2\right).
\end{equation}Therefore, the complexity of one inner iteration is
\begin{equation}
\mathcal{O}\!\left(
K M_r^3
+
M_t^3
+
\mathcal{C}_{\rm ch}
+
\mathcal{C}_{\rm NN}
+
\mathcal{C}_{\rm v}
\right),
\end{equation}
and the per-epoch complexity is
\begin{equation}
\mathcal{O}\!\left(
N_oN_i
\left(
K M_r^3
+
M_t^3
+
\mathcal{C}_{\rm ch}
+
\mathcal{C}_{\rm NN}
+
\mathcal{C}_{\rm v}
\right)
\right).
\end{equation}
The final trajectory evaluation and the Adam update of the meta-optimizer
parameters have the same order as the corresponding loss and gradient
computations and do not change the dominant scaling.

For large RIS arrays, the dominant RIS-dependent terms are channel
recomputation, meta-optimizer inference for $\boldsymbol{\theta}$ and
$\mathbf{R}$, and the pairwise spacing evaluation. Hence, the dominant
per-epoch RIS scaling is
\begin{equation}
\mathcal{O}\!\left(
N_oN_i
\left(
NL+HN+N^2
\right)
\right).
\end{equation}
Unlike conventional AO/SCA-based methods, the proposed H-GML framework
avoids repeated convex approximations and subproblem resolves for the
RIS phase and movable-position blocks, which reduce the practical
runtime for large ME-RIS systems.

\section{Numerical Results}

This section evaluates the performance of the proposed H-GML framework for
secure energy-efficient uplink transmission in an FD ME-RIS-assisted MA-BS
system. The BS is located at $(0,0,15)$ m, while the ME-RIS is deployed at
$(40,40,10)$ m. The users and Eves are located at a height $1.5$ m. The users
are randomly distributed around the RIS, whereas the Eves are randomly
distributed around the BS. A minimum separation of $10$~m is imposed between
users and between Eves. Moreover, the user--Eve distance is constrained to
lie in the interval $[40,100]$ m. Unless otherwise stated, each point is
averaged over the 100 independent channel realizations, and the same
set of realizations is used for all compared schemes in each sweep. The main
simulation parameters are summarized in Table~\ref{tab:sim_params}. 

\begin{table}[!t]
\scriptsize
\renewcommand{\arraystretch}{1.2}
\caption{Simulation Parameters}
\label{tab:sim_params}
\centering
\begin{tabular}{|l|c|l|c|}
\hline
\textbf{Parameter} & \textbf{Value} & \textbf{Parameter} & \textbf{Value} \\
\hline
$f_c$ & $3$\,GHz 
& $\eta$ & $10^{-9}$ \\
\hline
$\sigma_u^2, \sigma_d^2$ & $-97$\,dBm 
& $P_c$ & $20$\,dBm \\
\hline
$M_r, M_t$ & $6$ 
& $P_k^{\max}$ & $23$\,dBm \\
\hline
$K, E$ & $3,2$ 
& $P_{\rm AN}^{\max}$ & $10$\,dBm \\
\hline
$L_B, L_U$ & $4$ 
& $R_k^{\rm th}$ & $1$\,bps/Hz \\
\hline
$\beta_0$ & $-30$\,dB 
& $\eta_p$ & $0.1$ \\
\hline
$\alpha_{\rm BR},\alpha_{\rm Rk},\alpha_{\rm Re}$ & $2,2.2,2.4$ 
& $\alpha_{\rm Bk},\alpha_{\rm Be},\alpha_{\rm k,e}$ & $3.6$ \\
\hline
\end{tabular}
\end{table}

For clarity, we use the following naming convention. ``ME'' denotes movable
RIS elements, ``FE'' denotes fixed RIS elements, ``MA'' denotes movable BS
antennas, and ``FA'' denotes fixed BS antennas. The following schemes are
used for comparison.

\begin{itemize}
\item \textbf{EE (ME/MA, no Eve):} The ME-RIS and MA-BS system
is optimized for energy efficiency without secrecy constraints and without
AN. This scheme is assumed as a non-secure upper reference.

\item \textbf{SEE (ME/MA, no Eve knowledge):} The ME-RIS and MA-BS system is
designed without Eve CSI and is then evaluated under the actual Eve
channels. 

\item \textbf{SEE (ME/MA, with AN):} This is the
proposed design, including ME-RIS and MA-BS with the AN, and being aware of the CSI of Eves.

\item \textbf{SEE (ME/MA, no AN):} This baseline uses the same
ME-RIS and MA-BS architecture as the proposed scheme, without AN
transmission. 

\item \textbf{SEE (FE/MA, with AN):} The BS antennas are movable, but the RIS elements are fixed, with AN transmission. 

\item \textbf{SEE (ME/FA, with AN):} The RIS elements are movable, but the BS antennas are fixed, with AN transmission.

\item \textbf{SEE (FE/FA, with AN):} Both the RIS elements and BS antennas are fixed, with AN transmission.

\item \textbf{SEE (FE/FA, no AN):} Both the RIS elements and BS antennas are fixed, without AN transmission.

\item \textbf{SEE (FE/FA, no Eve knowledge):} Both the RIS elements and BS antennas are fixed, and the design is performed without Eve CSI.

\item \textbf{SEE (FE/FA, no Eve knowledge, random RIS):} The RIS phases are randomly selected, both the RIS elements and BS antennas are fixed, and Eve CSI is not used in the design.
\end{itemize}

\begin{figure}[htbp!]
    \centering
    \includegraphics[width=0.75\linewidth]{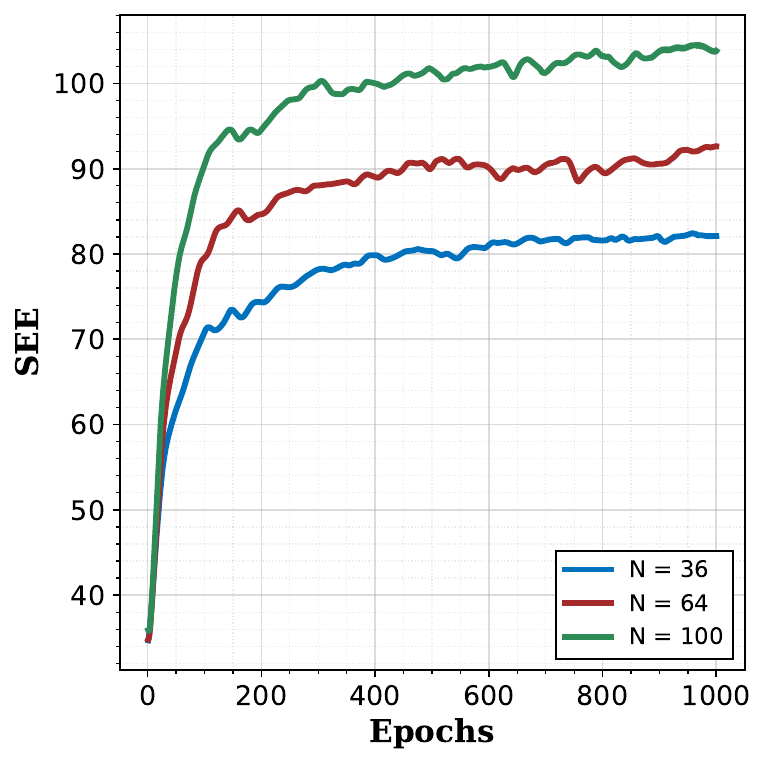}
    \caption{Convergence behavior of the proposed H-GML framework.}
    \label{fig:conv}
\end{figure}

Fig.~\ref{fig:conv} illustrates the convergence behavior of the proposed
H-GML framework for different numbers of RIS elements. The SEE rapidly increases during the initial
iterations and then stabilizes, which confirms the convergence of the
proposed H-GML optimizer. The final SEE improves as $N$ increases,
because a larger RIS provides higher passive beamforming gain and more
spatial degrees of freedom for strengthening the legitimate links while
suppressing information leakage.

\begin{figure}[!t]
    \centering
    \includegraphics[width=0.75\linewidth]{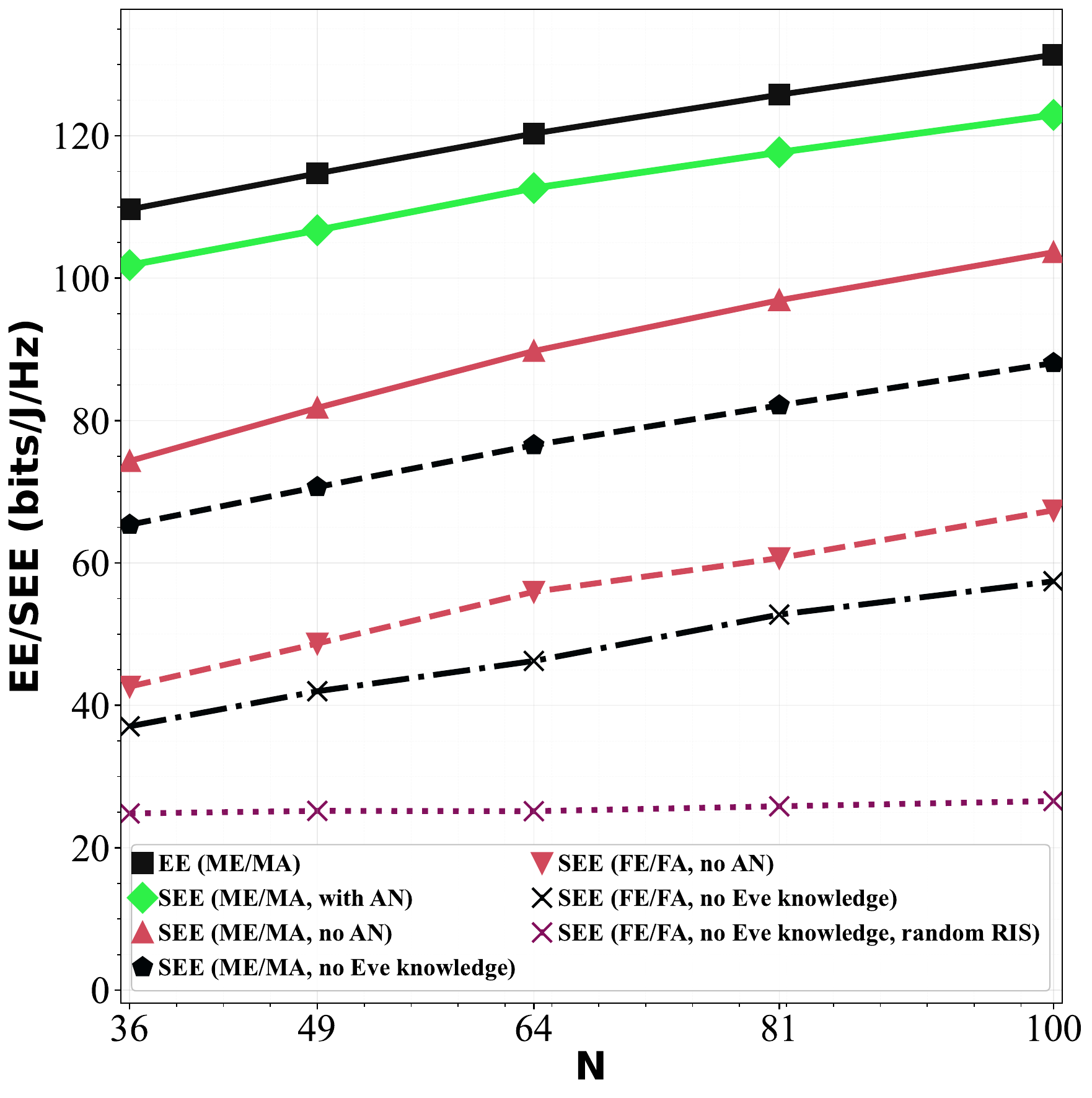}
    \caption{EE/SEE versus the number of RIS elements $N$.}
    \label{fig:vsN1}
\end{figure}
Fig.~\ref{fig:vsN1} illustrates the EE/SEE performance versus the number of RIS
elements $N$. As expected, the \textit{EE (ME/MA)} curve provides the highest
performance since it does not account for secrecy degradation or AN power
consumption. Among the secure schemes, \textit{SEE (ME/MA, with AN)} achieves
the best performance over the whole range of $N$ and increases steadily as $N$
grows, showing the benefit of additional RIS elements in improving
passive beamforming gain and spatial reconfigurability.
The proposed \textit{SEE (ME/MA, with AN)} scheme also provides a clear gain
over its main secure baselines. At $N=100$, it improves the SEE by about
$19\%$ compared with \textit{SEE (ME/MA, no AN)}, confirming the importance of
optimized AN in suppressing the cooperative Eves. Moreover, it achieves about
$40\%$ higher SEE than \textit{SEE (ME/MA, no Eve knowledge)}, which highlights
the role of Eve CSI in jointly shaping the AN, RIS phases, and movable antenna
positions. Compared with the baseline
\textit{SEE (FE/FA, no AN)}, the proposed scheme provides about $81\%$ gain at
$N=100$, verifying the advantage of jointly exploiting the BS-antenna and
RIS-element mobility, and AN. Finally, the nearly flat performance of
\textit{SEE (FE/FA, no Eve knowledge, random RIS)} confirms that random RIS
phase shifts cannot effectively support secure energy-efficient transmission.

\begin{figure}[!t]
    \centering
    \includegraphics[width=0.75\linewidth]{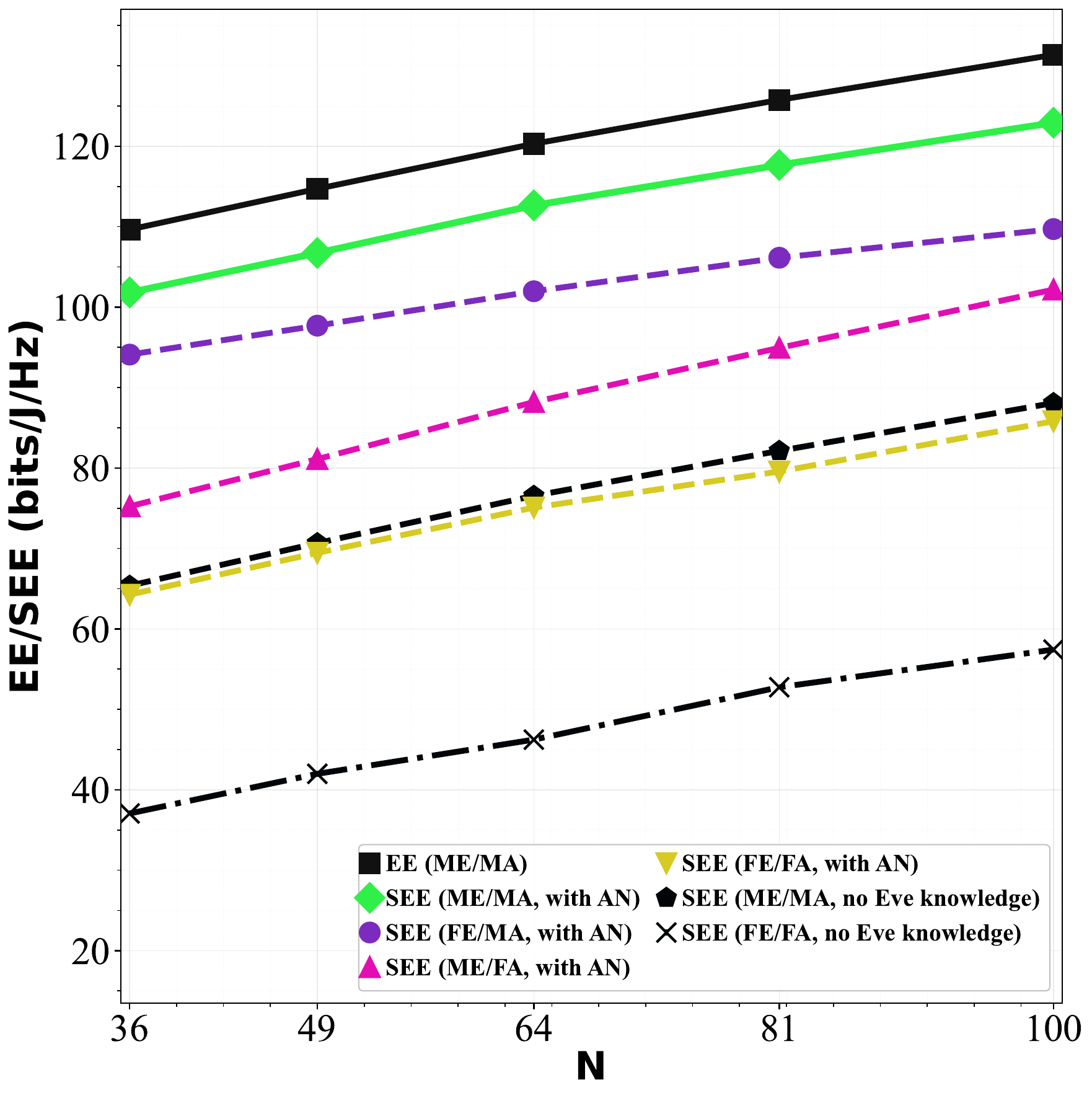}
    \caption{EE/SEE versus $N$ for different RIS and BS mobility scenarios.}
    \label{fig:vsN2}
\end{figure}

Fig.~\ref{fig:vsN2} compares the impact of RIS-element mobility and BS-antenna
mobility on the EE/SEE performance. The \textit{EE (ME/MA)} curve provides the
largest value, as it does not account for secrecy loss. Among the secure
schemes, \textit{SEE (ME/MA, with AN)} achieves the highest SEE for all values
of $N$, confirming that joint RIS-element and BS-antenna mobility provides the
largest spatial reconfiguration gain.
Among the partial-mobility schemes, \textit{SEE (FE/MA, with AN)} performs
better than \textit{SEE (ME/FA, with AN)}, indicating that BS-antenna mobility
is particularly effective for enhancing the receive-side beamforming and
SI and eavesdropping suppression. At $N=100$, the proposed
\textit{SEE (ME/MA, with AN)} scheme improves the SEE by about $13\%$ and
$21\%$ compared with \textit{SEE (FE/MA, with AN)} and
\textit{SEE (ME/FA, with AN)}, respectively. Moreover, compared with the fully
fixed \textit{SEE (FE/FA, with AN)} baseline, the proposed scheme provides
about $43\%$ SEE gain at $N=100$, which confirms the benefits of
jointly optimizing the BS-antenna positions and RIS-element locations.

The no-Eve-knowledge curves remain considerably below their Eve-aware
counterparts. In particular, at $N=100$, \textit{SEE (ME/MA, with AN)}
outperforms \textit{SEE (ME/MA, no Eve knowledge)} by about $40\%$, highlighting the importance of Eve CSI for jointly designing the AN direction, RIS phases,
and movable array geometries.

\begin{figure}[!t]
    \centering
    \includegraphics[width=0.75\linewidth]{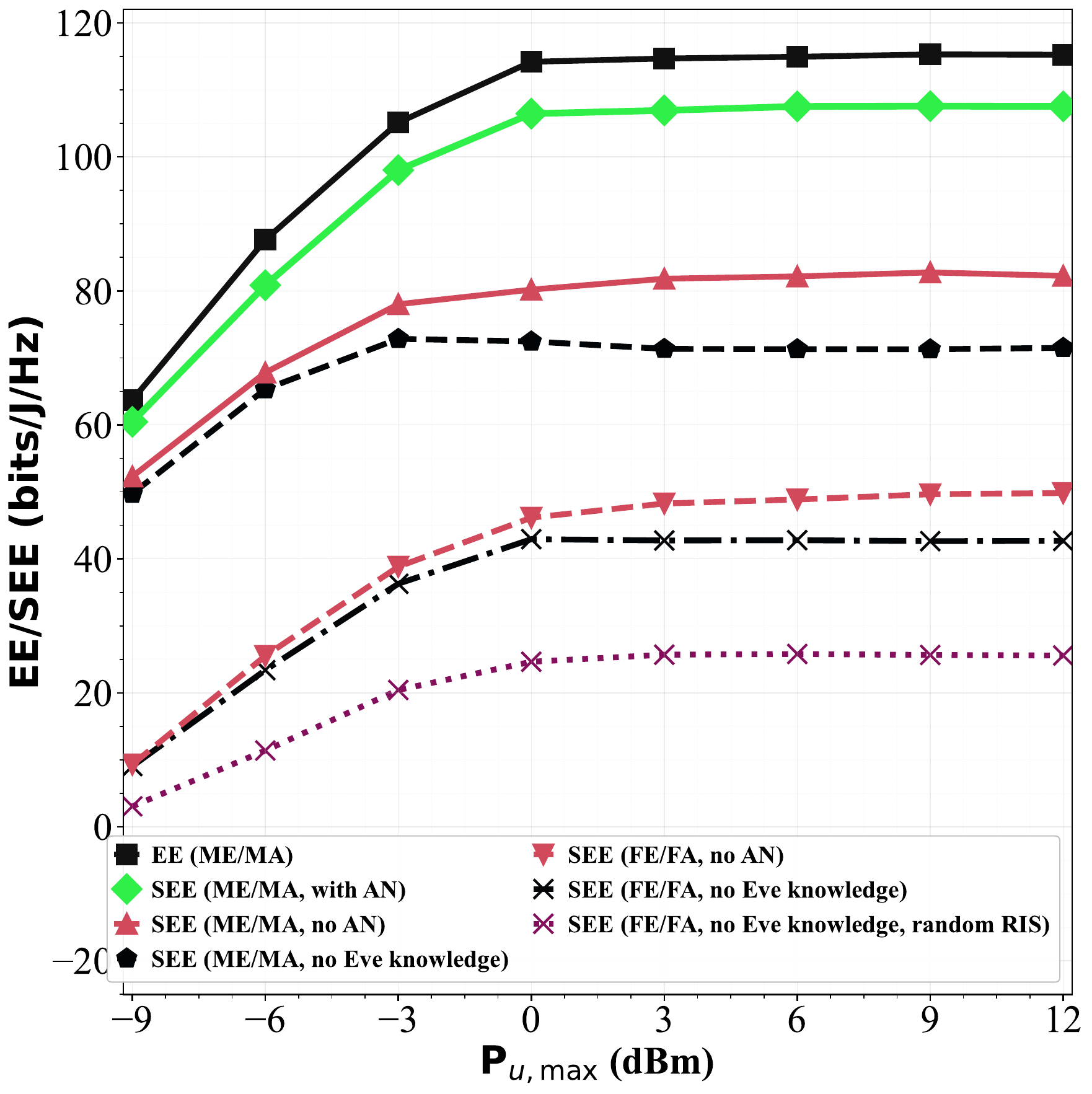}
    \caption{EE/SEE versus the maximum user transmit power $P_{u,\max}$.}
    \label{fig:vsP}
\end{figure}

Fig.~\ref{fig:vsP} shows the EE/SEE performance versus the maximum user
transmit power $P_{u,\max}$. For all schemes, the EE/SEE increases rapidly at
low transmit-power budgets due to the improved received SINR, and then
gradually saturates as the additional secrecy-rate gain becomes limited while the power consumption continues to increase.
Among the secure schemes, \textit{SEE (ME/MA, with AN)} achieves the highest
SEE over the entire range of $P_{u,\max}$. At $P_{u,\max}=12$ dBm, it provides
about $30\%$ and $50\%$ higher SEE than \textit{SEE (ME/MA, no AN)} and
\textit{SEE (ME/MA, no Eve knowledge)}, respectively, confirming the importance
of optimized AN and Eve-aware spatial design. The \textit{EE (ME/MA)} curve
remains above all secure schemes since secrecy loss and AN power consumption
are not considered.

The fixed-geometry baselines saturate at much lower values. In particular, the
proposed \textit{SEE (ME/MA, with AN)} scheme more than doubles the SEE of
\textit{SEE (FE/FA, no AN)} at 12 dBm, showing the benefit of
joint BS-antenna and RIS-element mobility. The low performance of the random
RIS baseline further shows that optimized RIS phase shifts are essential for
secure energy-efficient transmission.

\begin{figure}[htbp!]
    \centering
    \includegraphics[width=0.75\linewidth]{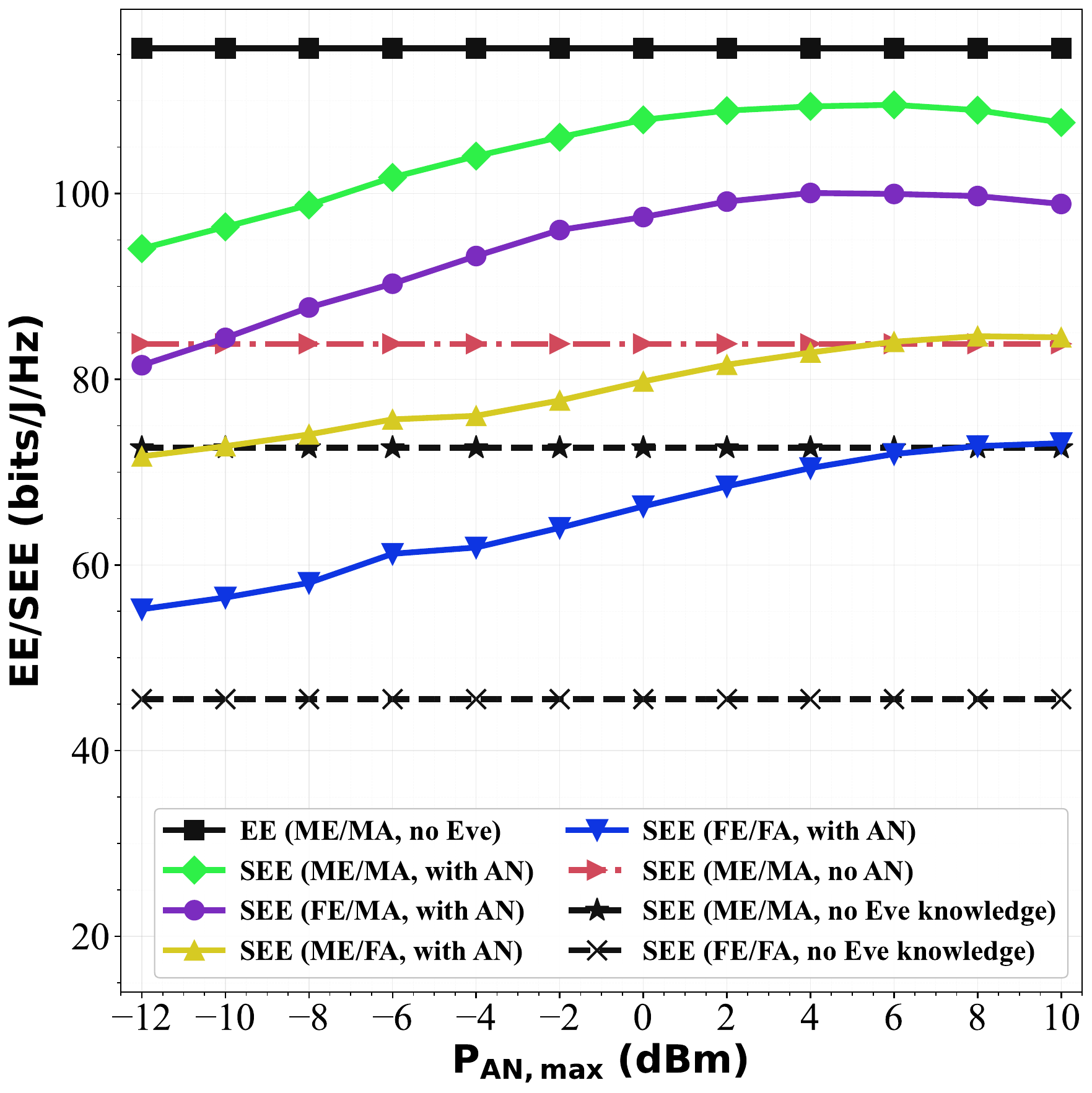}
    \caption{EE/SEE versus the maximum AN transmit power $P_{\rm AN,\max}$.}
    \label{fig:vsPAN}
\end{figure}

Fig.~\ref{fig:vsPAN} shows the EE/SEE performance versus the maximum AN power
$P_{\rm AN,\max}$. The \textit{EE (ME/MA, no Eve)} curve remains constant
because it corresponds to the no-Eve case and does not depend on the
AN budget. Similarly, the no-AN and no-Eve-knowledge baselines are almost
flat, since they cannot effectively exploit the available AN power.
In contrast, the AN-enabled schemes improve as $P_{\rm AN,\max}$ increases,
because a larger AN budget allows stronger interference to be directed toward
the Eves. Among them, \textit{SEE (ME/MA, with AN)} achieves the highest SEE
over the whole range, confirming the benefit of jointly optimizing AN, antennas/elements positions. At
$P_{\rm AN,\max}=10$ dBm, it provides about $29\%$ higher SEE than
\textit{SEE (ME/MA, no AN)}, which shows the importance of AN for
secrecy enhancement. Moreover, it outperforms \textit{SEE (FE/MA, with AN)}
and \textit{SEE (FE/FA, with AN)} by about $10\%$ and $47\%$, respectively,
highlighting the additional gain achieved by joint BS and RIS 
mobility. The gradual saturation at high AN budgets indicates that, beyond a
certain point, the secrecy-rate improvement becomes limited while the consumed
power continues to increase.

\begin{figure}[htbp!]
    \centering
    \includegraphics[width=0.75\linewidth]{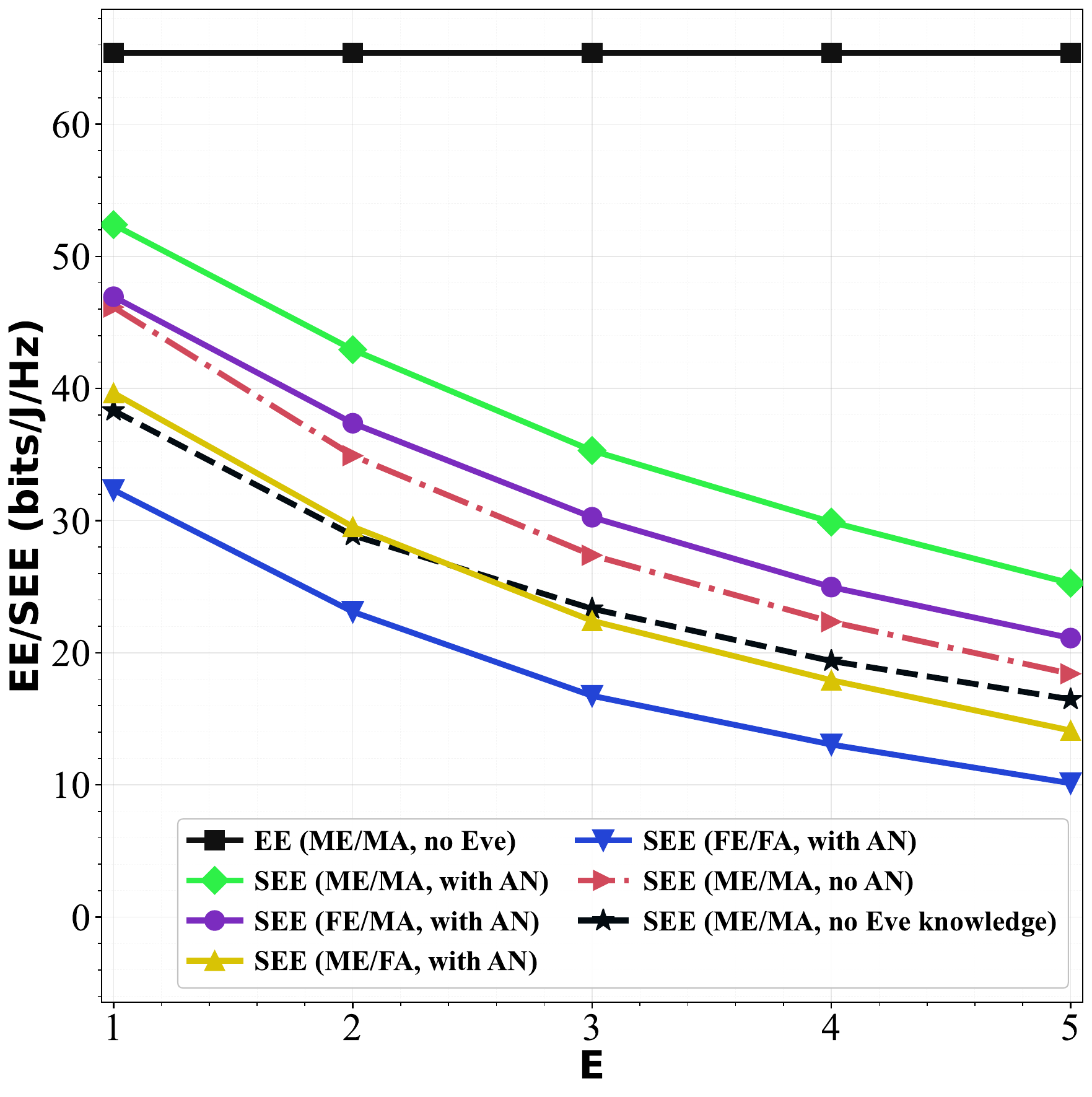}
    \caption{EE/SEE versus the number of cooperative Eves $E$.}
    \label{fig:vsE}
\end{figure}

Fig.~\ref{fig:vsE} shows the EE/SEE performance versus the number of
cooperative Eves $E$. The \textit{EE (ME/MA, no Eve)} curve remains constant
because it does not include secrecy in its objective. In contrast, all SEE
curves decrease as $E$ increases, since additional cooperative Eves increase
the overall eavesdropping capability and reduce the secrecy rate.

The \textit{SEE (ME/MA, with AN)} curve consistently achieves the highest
SEE among all secure schemes for every value of $E$. Although
\textit{SEE (ME/MA, with AN)} decreases as the number of Eves increases, it
maintains a clear advantage over \textit{SEE (ME/MA, no AN)},
\textit{SEE (FE/MA, with AN)}, \textit{SEE (ME/FA, with AN)},
\textit{SEE (FE/FA, with AN)}, and
\textit{SEE (ME/MA, no Eve knowledge)}. This robustness comes from the
joint optimization of AN and spatial reconfiguration: as the eavesdropping
environment becomes more severe, the optimizer can adjust the RIS elements,
BS antenna locations, AN direction, and transmit powers to better suppress
the Eve links.

The \textit{SEE (ME/MA, no AN)} curve suffers a stronger degradation as
$E$ increases, showing that passive beamforming and mobility alone are
insufficient in dense eavesdropping scenarios. The
\textit{SEE (FE/MA, with AN)} and \textit{SEE (ME/FA, with AN)} curves
remain below \textit{SEE (ME/MA, with AN)} because only one side of the
channel geometry is reconfigurable. The \textit{SEE (FE/FA, with AN)} curve
shows the largest loss among the AN-enabled schemes, indicating that
optimized AN is less effective when the system cannot reshape the
propagation geometry. Finally, \textit{SEE (ME/MA, no Eve knowledge)}
decreases significantly with $E$, confirming that Eve-aware optimization
becomes increasingly important as the number of cooperative Eves grows.

\begin{figure}
    \centering
    \includegraphics[width=0.85\linewidth]{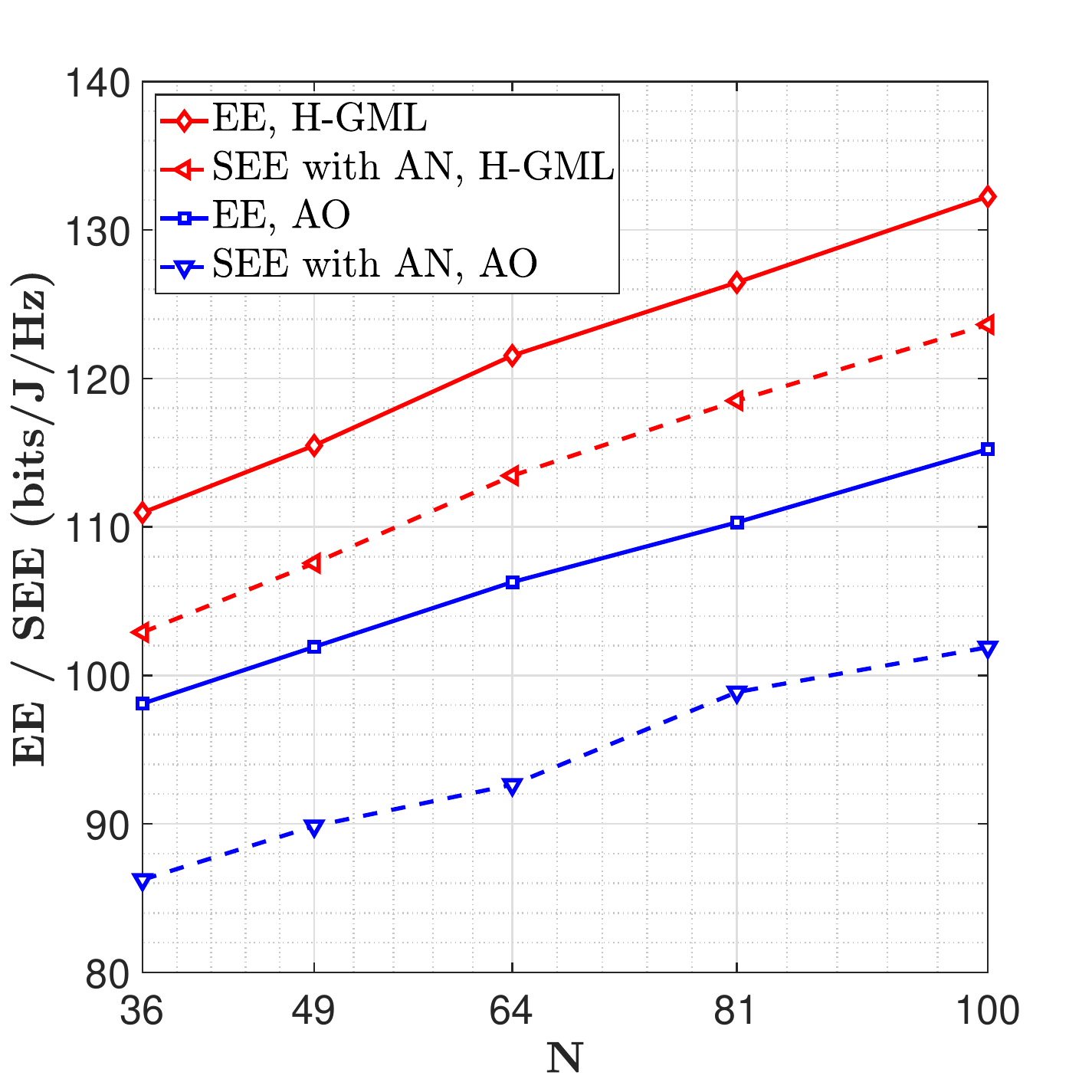}
    \caption{Performance comparison of H-GML and AO versus the N values.}
    \label{fig:optimality}
\end{figure}

Fig.~\ref{fig:optimality} compares the proposed H-GML framework with the model-based AO benchmark versus the number of RIS elements $N$. For both EE and SEE, the performance increases monotonically with $N$, since a larger ME-RIS provides higher passive beamforming gain and more spatial degrees of freedom for improving the legitimate uplink links and mitigating information leakage toward the Eves. As expected, the EE curves remain above the SEE curves because EE optimization does not account for secrecy loss or AN power consumption.
More importantly, H-GML consistently outperforms AO for both objectives over the whole range of $N$. At $N=100$, H-GML improves the EE by about $15\%$ and
the SEE by about $22\%$ compared with AO. The SEE gain is slightly larger, indicating that H-GML is particularly effective in handling the stronger coupling among power allocation, AN design, RIS phase shifts, and spatial mobility under secrecy-aware optimization. This confirms the advantage of the H-GML structure, which combines closed-form updates for the receive postcoders and AN direction with learned gradient-based updates for the remaining variables.

\section{Conclusion}

This paper studied secure energy-efficient uplink transmission in a
full-duplex ME-RIS-assisted MA-BS system with multiple cooperative
eavesdroppers. An Eve-aware H-GML framework was proposed to maximize the SEE
by jointly optimizing the transmit powers, BS receive postcoders, AN design, RIS phase shifts, and the positions of the BS antennas and RIS elements. By combining closed-form updates with neural meta-optimizers, the proposed framework provides an efficient solution for the resulting highly nonconvex joint optimization problem.

Numerical results showed that the proposed \textit{SEE (ME/MA, with AN)}
scheme consistently outperforms the considered secure baselines under different RIS sizes, user power budgets, AN power budgets, and numbers of Eves. The results confirmed the benefits of jointly exploiting BS antenna mobility, RIS element mobility, Eve-aware optimization, and AN transmission.
In particular, spatial mobility improves the channel geometry, while AN provides an effective mechanism for suppressing cooperative eavesdropping.

Overall, the proposed ME-RIS-assisted MA-BS design offers an effective
approach for improving both energy efficiency and physical-layer security in
future uplink networks. Future work may consider near-field propagation,
imperfect CSI, and online adaptation under user mobility.

\appendices
\section{Model-Based AO Solver for SEE Maximization}
\label{app:ao_solver}

This appendix summarizes the model-based AO
solver used as a deterministic benchmark for the proposed meta-learning
framework. The AO solver directly tackles the non-convex SEE maximization problem by sequentially updating the receive
postcoders, user transmit powers, AN design, RIS phase
shifts, and the movable positions of the BS antennas and RIS elements.
The solution relies on fractional programming, SCA, generalized Rayleigh quotient optimization, and
manifold-based phase optimization~\cite{rayleigh1,dinkelbach1967,Absil2008,Nocedal2006}.

The optimization variables are collected as
\begin{equation}
    \mathcal X \triangleq
    \left\{
    \{p_k\}_{k=1}^{K},\{\mathbf v_k\}_{k=1}^{K},
    \mathbf z,p_{\mathrm{an}},\boldsymbol\Phi,\mathbf T_r,\mathbf T_t,\mathbf R
    \right\},
\end{equation}
where $p_k$ is the transmit power of user $k$, $\mathbf v_k$ is the BS
receive postcoder, $\mathbf z$ is the normalized AN beamforming vector,
$p_{\mathrm{an}}$ is the AN power, $\boldsymbol\Phi$ is the RIS phase-shift
matrix, and $\mathbf T_r$, $\mathbf T_t$, and $\mathbf R$ denote the
locations of the BS receive antennas, BS transmit antennas, and RIS
elements, respectively. The SEE objective is
\begin{equation}
    \mathrm{SEE} =
    \frac{\sum_{k=1}^{K} [R_k-R_{E,k}]^{+}}
    {\frac{1}{\zeta_p}\left(\sum_{k=1}^{K}p_k+p_{\mathrm{an}}\right)+P_c},
    \label{eq:app_see}
\end{equation}
subject to the QoS constraints $R_k\ge R_k^{\mathrm{th}}$, power constraints,
unit-norm constraints on $\mathbf v_k$ and $\mathbf z$, unit-modulus RIS
constraints, movement-region constraints, and minimum-distance constraints.

For fixed $\{p_k\}$, $\mathbf z$, $p_{\mathrm{an}}$, $\boldsymbol\Phi$,
$\mathbf T_r$, $\mathbf T_t$, and $\mathbf R$, the receive postcoder of user
$k$ is obtained from the generalized Rayleigh quotient~\cite{rayleigh1,rayleigh2,ding2024secure}.
Let $\bar{\mathbf h}_k=\mathbf h_k+\mathbf G\boldsymbol\Phi\mathbf g_k$ and
\begin{equation}
    \mathbf R_k =
    \sigma_u^2\mathbf I
    +\eta p_{\mathrm{an}}(\mathbf F\mathbf z)(\mathbf F\mathbf z)^H
    +\sum_{j\neq k}p_j\bar{\mathbf h}_j\bar{\mathbf h}_j^H .
\end{equation}
Then,
\begin{equation}
    \mathbf v_k^{\star}
    =
    \frac{\mathbf R_k^{-1}\bar{\mathbf h}_k}
    {\left\|\mathbf R_k^{-1}\bar{\mathbf h}_k\right\|_2}.
    \label{eq:app_vk}
\end{equation}

For fixed postcoders and spatial variables, the user powers are updated by
Dinkelbach fractional programming combined with SCA~\cite{dinkelbach1967,boyd}.
Introducing $t_k\ge0$ to represent the positive secrecy contribution of user
$k$, the SCA subproblem at iteration $i$ is
\begin{subequations}
\small
\label{eq:app_power_sca}
\begin{align}
\max_{\{p_k\},\{t_k\}}~~&
\sum_{k=1}^{K}t_k
-\lambda\!\left(
\frac{\sum_{k=1}^{K}p_k+p_{\mathrm{an}}}{\zeta_p}+P_c
\right)
-\rho\|\mathbf p-\mathbf p^{(i)}\|_2^2
\\
\mathrm{s.t.}~~&
t_k\le
\widetilde R_k(\mathbf p)-\widetilde R_{E,k}(\mathbf p),
\quad \forall k,
\\
&
\widetilde R_k(\mathbf p)\ge R_k^{\mathrm{th}},
\quad \forall k,
\\
&
0\le p_k\le P_k^{\max},\quad t_k\ge0,
\quad \forall k,
\\
&
\|\mathbf p-\mathbf p^{(i)}\|_2\le \Delta^{(i)} .
\end{align}
\end{subequations}
Here, $\widetilde R_k(\mathbf p)$ is the concave lower bound of the legitimate
rate obtained by linearizing the interference logarithm, while
$\widetilde R_{E,k}(\mathbf p)$ is the affine upper approximation of the
cooperative eavesdropping rate. The trust-region and proximal terms improve
the accuracy of the local approximation and stabilize the SCA iterations
~\cite{trust1,trust2,trust3}. The resulting convex program can be solved
using standard convex optimization tools such as CVX~\cite{cvx}. After each
SCA update, the exact rates are recomputed and the Dinkelbach parameter is
updated as
\begin{equation}
    \lambda =
    \frac{\sum_{k=1}^{K}[R_k(\mathbf p)-R_{E,k}(\mathbf p)]^+}
    {\frac{1}{\zeta_p}\left(\sum_{k=1}^{K}p_k+p_{\mathrm{an}}\right)+P_c}.
\end{equation}

The AN direction is designed to increase AN leakage toward the cooperative
Eves while suppressing residual self-interference at the BS. Thus,
$\mathbf z$ is obtained from the generalized Rayleigh quotient
\cite{rayleigh1,golub}
\begin{equation}
    \max_{\|\mathbf z\|_2=1}
    \frac{\mathbf z^H\mathbf A\mathbf z}
    {\mathbf z^H(\mathbf B+\epsilon_z\mathbf I)\mathbf z},
    \label{eq:app_z}
\end{equation}
where
\begin{equation}
    \mathbf A=\sum_{e=1}^{E}\mathbf f_e^H\mathbf f_e,\qquad
    \mathbf B=\eta\sum_{k=1}^{K}\mathbf F^H\mathbf v_k\mathbf v_k^H\mathbf F .
\end{equation}
The solution is the dominant generalized eigenvector of
$(\mathbf A,\mathbf B+\epsilon_z\mathbf I)$. With $\mathbf z$ fixed,
$p_{\mathrm{an}}$ is optimized over $[0,P_{\mathrm{an}}^{\max}]$ using the
scalar Dinkelbach problem~\cite{dinkelbach1967}
\begin{equation}
    \max_{0\le p_{\mathrm{an}}\le P_{\mathrm{an}}^{\max}}
    \mathrm{SSR}(p_{\mathrm{an}})
    -\lambda
    \left(
    \frac{\sum_{k=1}^{K}p_k+p_{\mathrm{an}}}{\zeta_p}+P_c
    \right).
    \label{eq:app_pan}
\end{equation}
The one-dimensional interval is partitioned according to the breakpoints
where $R_k(p_{\mathrm{an}})=R_{E,k}(p_{\mathrm{an}})$. Within each interval,
the active user set is fixed; candidate stationary points are found by
bisection on the derivative, and the best exact objective value is selected.

For the RIS phase shifts, define
$\boldsymbol\theta=[e^{j\vartheta_1},\ldots,e^{j\vartheta_N}]^T$ and
$\boldsymbol\Phi=\mathrm{diag}(\boldsymbol\theta)$. The unit-modulus
constraint is handled on the torus manifold~\cite{Absil2008,AlaaEldin2022}
by maximizing
\begin{equation}
    J(\boldsymbol\vartheta)
    =
    \mathrm{SSR}(\boldsymbol\vartheta)
    -\zeta\sum_{k=1}^{K}
    \left[ R_k^{\mathrm{th}}-R_k(\boldsymbol\vartheta)\right]_+^2 .
    \label{eq:app_phi_obj}
\end{equation}
The gradient is approximated by central finite differences~\cite{Nocedal2006},
\begin{equation}
    [\mathbf q^{(i)}]_n =
    \frac{
    J(\boldsymbol\vartheta^{(i)}+\Delta_\vartheta\mathbf e_n)
    -
    J(\boldsymbol\vartheta^{(i)}-\Delta_\vartheta\mathbf e_n)
    }
    {2\Delta_\vartheta},
\end{equation}
and an Armijo backtracking line search is used to select $\alpha^{(i)}$
~\cite{Armijo1966}. The phase vector is then retracted as
\begin{equation}
    \boldsymbol\vartheta^{(i+1)}
    =
    \mathrm{mod}\!\left(
    \boldsymbol\vartheta^{(i)}+\alpha^{(i)}\mathbf q^{(i)},2\pi
    \right).
\end{equation}

Finally, the movable positions $\mathbf T_t$, $\mathbf T_r$, and $\mathbf R$
are updated sequentially using trust-region SCA~\cite{Nocedal2006,trust1,trust2,trust3}.
For a generic position block
$\mathbf X\in\{\mathbf T_t,\mathbf T_r,\mathbf R\}$, define the secrecy gap
at the current iterate as
\begin{equation}
    \Upsilon_k^{(i)}=R_k^{(i)}-R_{E,k}^{(i)} .
\end{equation}
For active users with $\Upsilon_k^{(i)}>0$, the secrecy gap is locally
approximated as
\begin{equation}
    \widetilde{\Upsilon}_k^{(i)}(\mathbf X)
    =
    \Upsilon_k^{(i)}
    +
    \mathrm{vec}\!\left(\nabla_{\mathbf X}(R_k-R_{E,k})\right)^T
    \mathrm{vec}\!\left(\mathbf X-\mathbf X^{(i)}\right).
\end{equation}
The convexified position subproblem is
\begin{subequations}
\label{eq:app_position_sca}
\begin{align}
\max_{\mathbf X,\{s_k\}}~~&
\sum_{k=1}^{K}s_k
\\
\mathrm{s.t.}~~&
\widetilde R_k^{(i)}(\mathbf X)\ge R_k^{\mathrm{th}},
\quad \forall k,
\\
&
0\le s_k\le \widetilde{\Upsilon}_k^{(i)}(\mathbf X),
\quad \forall k:\Upsilon_k^{(i)}>0,
\\
&
s_k=0,\quad \forall k:\Upsilon_k^{(i)}\le0,
\\
&
\mathbf x_q\in\mathcal A,\quad \forall q,
\\
&
\frac{(\mathbf x_q^{(i)}-\mathbf x_\ell^{(i)})^T}
{\|\mathbf x_q^{(i)}-\mathbf x_\ell^{(i)}\|_2}
(\mathbf x_q-\mathbf x_\ell)\ge d_0,\quad \forall q\neq \ell,
\\
&
\|\mathbf x_q-\mathbf x_q^{(i)}\|_2\le \Delta^{(i)},\quad \forall q,
\end{align}
\end{subequations}
where $\mathbf x_q$ denotes the corresponding antenna or RIS-element
coordinate and $\mathcal A$ is the associated feasible movement region.
This convexified problem can be solved with CVX~\cite{cvx}.

\begin{algorithm}[!t]
\caption{AO-Based SEE Maximization Algorithm}
\label{alg:app_ao}
\begin{algorithmic}[1]
\State \textbf{Input:} Feasible initial point $\mathcal X^{(0)}$, tolerance
$\epsilon$, and maximum AO iterations $I_{\max}$.
\State Compute $\mathrm{SEE}^{(0)}$ and set $i=0$.
\Repeat
    \State Set $i\leftarrow i+1$.
    \State Update $\{\mathbf v_k^{(i)}\}$ using \eqref{eq:app_vk}.
    \State Update $\mathbf p^{(i)}$ by solving \eqref{eq:app_power_sca}.
    \State Update $\mathbf z^{(i)}$ using \eqref{eq:app_z}.
    \State Update $p_{\mathrm{an}}^{(i)}$ using the scalar Dinkelbach search
    in \eqref{eq:app_pan}.
    \State Update $\boldsymbol\Phi^{(i)}$ by maximizing
    \eqref{eq:app_phi_obj} on the unit-modulus manifold.
    \State Sequentially update $\mathbf T_t^{(i)}$, $\mathbf T_r^{(i)}$,
    and $\mathbf R^{(i)}$ using \eqref{eq:app_position_sca}.
    \State Compute $\mathrm{SEE}^{(i)}$.
\Until{$\frac{|\mathrm{SEE}^{(i)}-\mathrm{SEE}^{(i-1)}|}
{\mathrm{SEE}^{(i-1)}}\le\epsilon$
or $i\ge I_{\max}$}
\State \textbf{Output:} $\mathcal X^\star=\mathcal X^{(i)}$.
\end{algorithmic}
\end{algorithm}

The accepted AO updates generate a non-decreasing SEE sequence. Therefore,
the model-based solver converges to a stationary local solution under the
standard assumptions of SCA and fractional programming~\cite{Nocedal2006,dinkelbach1967,boyd}.
In the proposed meta-learning framework, this AO solver is used as a
high-quality benchmark, while meta-learning reduces the need for repeatedly
solving the full iterative optimization problem for new channel realizations.

\bibliographystyle{IEEEtran}
\bibliography{meta_learning/references_meta.bib}

\vfill

\end{document}